\documentclass[apj]{emulateapj}

\newcommand{\ha}{H$\alpha$}
\newcommand{\hb}{H$\beta$}
\newcommand{\kms}{ \ifmmode{\rm km\thinspace s^{-1}}\else km\thinspace s$^{-1}$\fi}
\newcommand{\dg}{\ifmmode{^{\circ}}\else $^{\circ}$\fi}
\newcommand{\vlsr}{\ifmmode{v_{\rm LSR}}\else $v_{\rm LSR}$\fi}
\newcommand{\iha}{\ifmmode{I_{\rm{H}\alpha}} \else $I_{\rm H \alpha}$\fi}
\newcommand{\ihb}{\ifmmode{I_{\rm{H}\beta}} \else $I_{\rm H \beta}$\fi}

\newcommand{\hi}{\ion{H}{1}}
\newcommand{\hii}{\ion{H}{2}}

\newcommand{\nh}{$N_{\rm{H}}$}
\newcommand{\lb}{\ifmmode{(\ell,b)} \else $(\ell,b)$\fi}
\newcommand{\nii}{[\ion{N}{2}]}
\newcommand{\sii}{[\ion{S}{2}]}
\newcommand{\oiii}{[\ion{O}{3}]}

\newcommand{\av}{\ifmmode{A(V)}\else $A(V)$\fi}
\newcommand{\ebv}{\ifmmode{E(B-V)}\else $E(B-V)$\fi}

\shorttitle{Low Extinction Window into Inner Galaxy}
\shortauthors{Madsen and Reynolds}

\begin{document}

\title{An Investigation of Diffuse Interstellar Gas toward a Large,
  Low Extinction Window into the Inner Galaxy}

\author{G. J. Madsen\altaffilmark{1} and R. J. Reynolds}
\affil{Department of Astronomy, University of Wisconsin--Madison,
475 N. Charter Street, Madison, WI 53706}
\altaffiltext{1}{Current address: Anglo-Australian Observatory, P.O.~Box 296, Epping, NSW 1710, Australia}
\email{madsen@aao.gov.au}

\slugcomment{Accepted to ApJ}

\begin{abstract}

\ha\ and \hb\ spectroscopy with the Wisconsin H-Alpha Mapper (WHAM) reveals a strong concentration of high velocity emission in a $\approx 5\dg \times
5\dg$ area centered near \lb\ $= (27\dg,-3\dg)$, known as the Scutum
Cloud. 
The high velocities imply that we are detecting optical emission from near
the plane of the Galaxy out to the tangent point at heliocentric
distances of $D_\odot \gtrsim 6$ kpc, assuming the gas participates in
circular Galactic rotation.   
The ratio of the \ha\ to \hb\ emission as a function of velocity
suggests that dust along these lines of sight   produces a total
visual extinction of $A_{V} \approx 3$ at $D_\odot \sim$ 6 kpc.
This makes it possible to use optical emission lines to explore the physical conditions of ionized gas in the inner Galaxy. At a Galactocentric distance $R_G \approx 4$ kpc, for example, we find that the H$^+$ has an rms midplane density of $\approx$ 1 cm$^{-3}$ with a vertical scale height of $\approx$ 300 pc.
We also find evidence for an increase in the flux of Lyman continuum
photons and an increase in the ratio of ionized to neutral hydrogen
toward the inner Galaxy. 
We have extended the measurements of \ebv\ in this direction to distances far beyond what has been accessible through stellar photometry and find \ebv/\nh\ to be near the local mean of 1.7 $\times$ 10$^{-22}$ cm$^{2}$ mag, with evidence for an increase in this ratio at $R_G \approx$ 4 kpc. 
Finally, our observations of \nii$~\lambda6583$, \sii$~\lambda6716$, and \oiii$~\lambda5007$ toward the window reveal that in the inner Galaxy the temperature of the gas and the ionization state of oxygen increase with increasing height from the midplane. 

\end{abstract}


\section{INTRODUCTION}

A thorough understanding of interstellar matter and processes in the
Galaxy requires a full exploration of all phases of the interstellar
medium (ISM) in a variety of environments,  ranging from the tenuous
halo to the energetic nucleus. 
Diffuse, warm ionized gas is one of the major phases of the ISM of our
Galaxy \citep[see reviews by][]{KH87, Reynolds91b, Mathis00,
  Ferriere01}. 
In the solar neighborhood, the warm ionized medium (WIM) consists of
regions of warm (10$^{4}$ K), low-density (10$^{-1}$ cm$^{-3}$),
nearly fully ionized hydrogen with a scale height of 1 kpc, and
appears to be present in every direction in the sky
\citep[e.g.,][]{Reynolds91a, NCT92, TC93, Haffner+03}. 
However, the origin and physical conditions within the WIM remain
poorly understood. 
Recent observational studies of the WIM in the solar neighborhood and
in the Perseus spiral arm of the Galaxy highlight some of the
problems, including how Lyman continuum photons are able to penetrate
the seemingly ubiquitous \hi\ to  ionize the large-scale WIM, and the
inability of photoionization models to explain some of the optical
line ratios \citep[e.g.,][]{HRT99, DM94, HW03}. 
Furthermore, the temperature and ionization conditions of diffuse
ionized gas in the {\emph{inner}} Galaxy are likely to differ
significantly from the local conditions, given its higher
star-formation rate, larger pressure, and stronger UV flux
\citep{Heiles+96a, Heiles+96b, RA01}.  
The presence of the inner Galactic molecular ring \citep{DHT01}, for
example, suggests that significant star formation is occurring
interior to the solar circle.  

Investigations of ionized gas through optical emission lines toward
the inner Galaxy, especially those regions close to the midplane where
the ionizing stars are located, were thought to be prohibitive because
interstellar dust severely obscures most of the optical light.   
As a result, studies of optical and ultraviolet emission sources near
the plane are largely constrained to the nearest few kiloparsecs. 
A potential alternative to optical studies of ionized gas in the inner
Galaxy is provided through radio recombination lines (RRLs), free-free
emission, and infrared emission, which penetrate the interstellar
dust.  
However, the current sensitivity of RRL and free-free surveys limits
these studies to only the brightest regions of ionized gas, with
emission measures greater than $\sim 500$\ cm$^{-6}$~pc
\citep{Heiles+96b}. In addition, diffuse emission from RRLs may need
to be corrected for high $n$-level effects in order to derive accurate
physical properties of the gas, which may be difficult to estimate
\citep{Afflerbach+96, BS72, Shaver80}.  Free-free continuum emission
contains no kinematic information, making it difficult to determine
the variations in the properties of the ionized gas along the line of
sight.   
Techniques for the detection and study of infrared emission lines from
diffuse ionized gas are under development and are currently limited to
emission from hydrogen \citep{Kutyrev+03}. 
Optical emission line studies, on the other hand, are sensitive to
much lower apparent emission measures, down to $\lesssim 0.02$\
cm$^{-6}$~pc \citep{Madsen+01, Weymann+01, Gallagher+03}, and often
include lines from a variety of ions from which physical properties of
the gas may be inferred. 
They also have the advantage of being relatively strong and providing
velocity information. The principal problem with optical lines is
their sensitivity to interstellar extinction. 

To overcome this shortcoming, we can take advantage of the highly
non-uniform distribution of interstellar dust, which allows us to see
further into the plane at some locations compared to others. Baade's
Window, for example, is a well known region of relatively low
extinction, about 1\dg\ in size, centered near $\lb = (1\dg,-4\dg)$
\citep{Baade63}.  With a total \av\ of only $\sim 1.5$ out to the
center of the Galaxy \citep{Stanek96}, this piece of sky has been the
focus of numerous UV, optical, and IR stellar photometric and
spectroscopic investigations used to infer the properties of the
Galactic bulge \citep[see review by][]{Frogel88}.  While its small
Galactic longitude makes Baade's window well suited for studies of the
bulge, the lack of a radial velocity spread toward this direction
makes emission line studies of the interstellar medium along the line
of sight problematic. 

\begin{figure}[htp]
\includegraphics[scale=0.8]{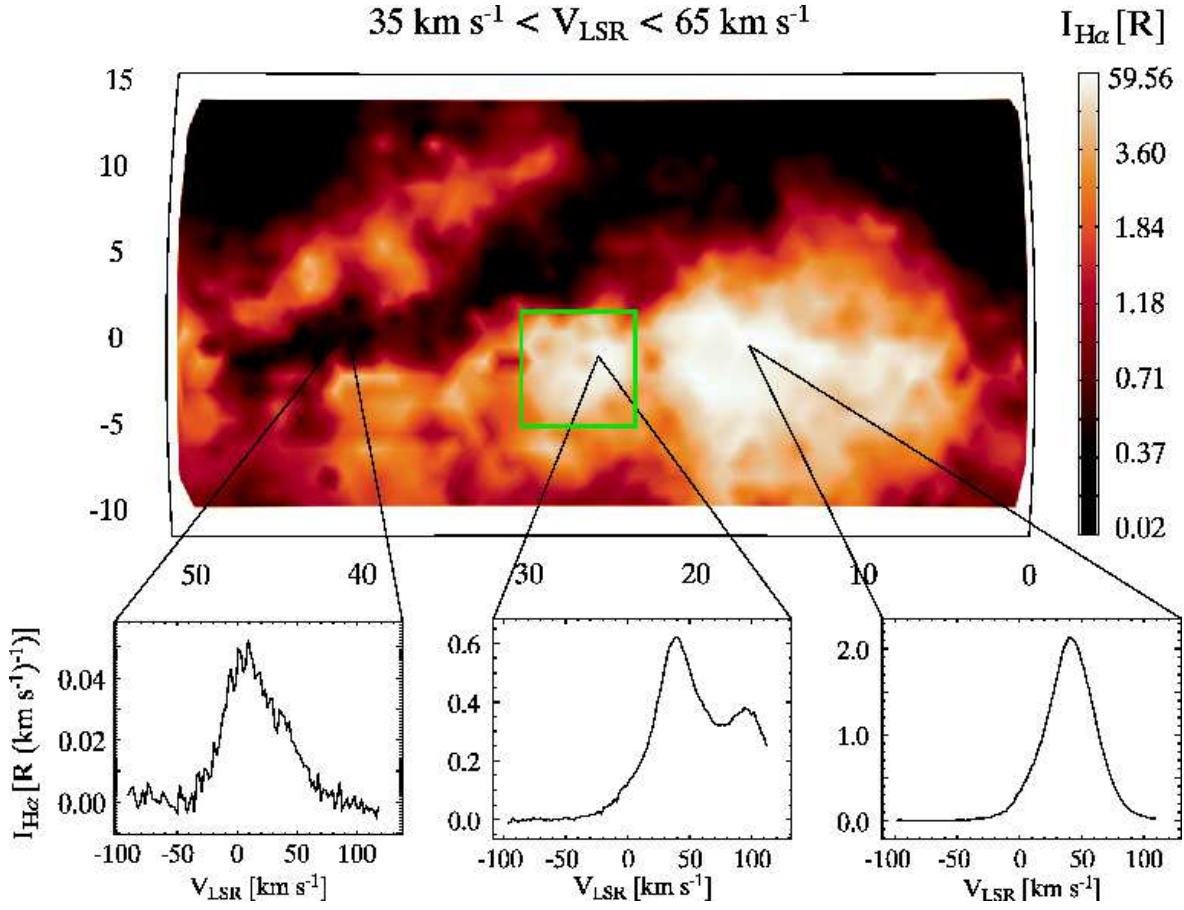}
\caption{\ha\ map and spectra from the WHAM-NSS toward the inner
  Galaxy. The large dark feature in the middle of the map is the
  Aquila rift, a nearby dust cloud (d $\approx$ 250 pc) that is
  obscuring the \ha\ emission behind it with $v_{\rm{LSR}} >$
  +25~\kms. The three diagrams below the map show the \ha\ spectra
  toward the indicated directions, revealing the change in the
  obscuration of emission from the local neighborhood (0~\kms), the
  Sagittarius arm (+50~\kms), and the Scutum arm (+80~\kms). Note the
  significant change in the intensity scale between these spectra. The
  green box is the low extinction window toward the `Scutum Star
  Cloud' with emission out to the tangent point
  velocity. \label{fig:1}} 
\end{figure}

Another much larger, and more suitable, window for ISM studies lies in
the direction of the so called `Scutum Star Cloud'.  This $\sim 5\dg$
diameter `cloud', centered near $\lb = (27\dg, -3\dg)$, is one of the
optically brightest regions of the Galaxy.  Its unusually high
apparent stellar density led early astronomers to believe it was a
region with an enhanced concentration of stars.  Subsequently, UV,
optical, and IR spectrophotometry of these stars have shown that this
area has unusually low extinction, with an \av\ $\approx 2$ out to a
distance of 2 kpc \citep{Albers62, KBF85, Reichen+90}. 
The presence of moderate and high velocity emission resulting from
differential Galactic rotation, and the observational imprint of dust
clouds toward this `cloud',  are evident in the recently completed
WHAM survey of the northern sky in \ha\ \citep[WHAM-NSS,
][]{Haffner+03}. 
Figure~\ref{fig:1} provides an \ha\ map from this survey showing the
emission at velocities \vlsr\ between +35 \kms\ and +65 \kms\ toward
the inner Galaxy. 
The \ha\ emission is given in units of Rayleighs\footnotemark
\footnotetext{$1R=10^6/4\pi$~photons~cm$^{-2}$~s$^{-1}$~str$^{-1}=2.4\times10^{-7}$~erg~cm$^{-2}$~s$^{-1}$~str$^{-1}$~at \ha}.
In this general direction, these velocities
correspond to emission from the Sagittarius spiral arm of the Milky
Way, which is at a distance of about 2--3 kpc \citep{TC93}. The large
dark feature extending from $\lb = (50\dg, 0\dg)$ to $\lb = (20\dg,
+10\dg)$ is the Aquila rift, a well-known, nearby  dust cloud
\citep[$d \approx 250$ pc,][]{SCB03}, which is obscuring the emission
behind it with \vlsr\ $\gtrsim$ +25 \kms.   

The three spectra underneath the emission line map in
Figure~\ref{fig:1} show more clearly the effect of extinction by
dust. The middle spectrum, toward the Scutum cloud, shows not only
emission from the local neighborhood (0 \kms) and the Sagittarius
spiral arm ($\approx +50\ \kms$), but the more distant emission from
the Scutum spiral arm ($\approx +75\ \kms$) and all the way out to the
tangent point velocity ($\approx  +100\ \kms$).  
The lack of detectable emission at some of these velocities in the
other two directions is believed to be caused by the inhomogeneous
distribution of interstellar dust.  
Assuming that ionized gas resides in the Sagittarius spiral arms at an
$l \approx 40\dg$, the leftmost spectrum shows only the weak emission
from nearby interstellar gas  in front of the Aquila rift.
The rightmost spectrum is in a direction which crosses both the
Sagittarius and Scutum arms, and shows only emission from local
sources and the Sagittarius arm, with dust extinguishing the emission
at larger distances. The green box in the emission map outlines the
low extinction window, which shows emission all the way out to the
tangent point velocity. 
Of course, the \ha\ data alone are not enough to confirm that this
picture of the distribution of dust along these lines of sight is
correct, or that we are detecting emission at large heliocentric
distances. The combination of the \ha\ and \hb\ spectra that extend
out beyond the tangent point velocity can verify that increased
extinction is indeed associated with the emission at increasingly
positive velocities. 

Here, we report on the detection of high velocity \ha\ and \hb\
emission from the warm, diffuse ionized gas toward the inner Galaxy,
including directions toward the Scutum Cloud, and we investigate the
nature of the diffuse ISM toward this unique window. 
We describe our spectroscopic observations in \S\ref{sec:obs}, with
particular emphasis on the importance of removing atmospheric lines
that contaminate the spectra.  
We present these results in \S\ref{sec:emission},  confirming the
presence of high velocity \ha\ and \hb\ emission and discussing their
relative strengths.  
A method for estimating the extinction toward this window, based on
the \ha\ and \hb\ spectra, is discussed in \S\ref{sec:ext}. In
\S\ref{sec:scale}, we correct the \ha\ emission for this extinction
and infer some intrinsic properties of the ionized gas, including its
density and scale height.  A comparison between the diffuse \ha\ and
\hi\ emission in this window is presented in \S\ref{sec:neutral},
including estimates of the relative column densities and flux of
ionizing photons in the inner Galaxy. We compare our extinction
measurements with the standard, empirical relationship between \ebv\
and $N_H$ along these lines of sight in \S\ref{sec:ebv}. We discuss
observations of \nii$~\lambda6583$, \sii$~\lambda6716$, and
\oiii$~\lambda5007$ toward a small number of directions in the window
and infer some basic physical properties of the ionized gas in
\S\ref{sec:ions}. We summarize our results in \S\ref{sec:summary}. 
 
\section{OBSERVATIONS}
\label{sec:obs}

All of the observations were carried out with the Wisconsin H-Alpha
Mapper (WHAM) spectrometer.  WHAM consists of a 0.6~m siderostat
coupled to a 15 cm dual-etalon Fabry-Perot system \citep{TuftePhD,
  Haffner+03} and produces an optical spectrum integrated over its
circular, $1\dg$ diameter field of view at a spectral resolution of 12
\kms\ within a 200 \kms\ wide spectral window that can be centered on
any wavelength between 4800 and 7400 \AA. WHAM was specifically
designed to detect very faint optical emission lines from the diffuse
interstellar medium.  It is located at the Kitt Peak National
Observatory in Arizona and is completely remotely operated from
Madison, Wisconsin. The WHAM Northern Sky Survey (WHAM-NSS), carried
out between 1997 January and 2000 April, has mapped the entire
northern sky ($\delta > -30\dg$) in \ha\ within $\approx \pm\ 100$
\kms\ of the local standard of rest \citep{Haffner+03}. 

One of the many intriguing results from the WHAM-NSS is the
significant amount of \ha\ emission that appears to extend beyond the
Survey's positive velocity limit ($\vlsr \approx +100~\kms$) in a
region with $33\dg < l < 20\dg, -5\dg < b < 0\dg$, particularly in the
general direction of the Scutum Cloud.  This prompted followup \ha\
observations obtained between 2002 May and 2002 August that extended
the original velocity coverage out to $\vlsr = +150$ \kms\ over an
area within $5\dg < l < 40\dg, -5\dg < b < +5\dg$.  Because these
velocities suggested that this emission may be arising from the inner
Galaxy, we also obtained \hb\ observations toward the same area over
the same velocity interval. The \hb\ data allows us to quantify the
total extinction and the variations of extinction with velocity along
the line of sight, and to verify that the emission is arising from
large heliocentric distances. 

These new observations were taken in the same manner as the original
Survey data \citep{Haffner+03}. 
The observed region was divided into eight observational `blocks',
with each block covering a $\approx 7\dg \times 7\dg$ area. 
Each direction was observed once in \ha\ and once in \hb\ for 30 sec
and 60 sec, respectively.  
Additionally, 13 individually pointed directions, those which showed
the strongest evidence of high-velocity \ha\ emission, were observed
with twice the above exposure times to obtain a higher signal-to-noise
ratio. These same 13 directions were also observed in the lines of
\nii, \sii, and \oiii. 

All of the spectra were contaminated by several weak atmospheric
emission lines that tarnish the interstellar Galactic spectra
\citep{Hausen+02-apj}. The identification and removal of these
terrestrial lines from the spectra is essential in order to compare
accurately the spectral profiles and relative strengths of the
components.  In particular, the \ha\ and \hb\ spectra are affected by
the relatively bright \ha\ and \hb\ geocoronal emission lines,
produced by the excitation of neutral hydrogen in the Earth's
atmosphere from scattered solar Ly$\beta$ and Ly$\gamma$ photons,
respectively. An additional challenge in the data reduction was the
difficulty in characterizing the continuum, because both interstellar
and terrestrial emissions are present almost everywhere across the 200
\kms\ spectral window. 
For each spectrum, a least-squares fit was performed, which consisted
of the the sum of a linear continuum, atmospheric lines, and the
Galactic emission. 
The determination of the best fit parameters for each of these
components of the spectrum, and the subsequent removal of the
contaminating atmospheric lines and the continuum, is outlined below
for the \ha, \hb, and other emission line data separately. A more
detailed discussion of the data reduction procedure of WHAM data can
be found in \citet{Haffner+03}.  

Another source of systematic uncertainty is the shape of the
underlying continuum due to stellar Fraunhofer lines that may be
present in diffuse starlight. This effect has yet to be studied
quantitatively.  
The large natural width of the strongest stellar lines, as well as the
kinematic broadening along a line of sight, reduces the magnitude of
this effect.  
We see no evidence for significant absorption in our spectra, and have
not attempted to make a correction for this effect. 

Because of our interest in the ratio of the \ha\ to \hb\ emission as a
function of velocity, calibrating the intensities of the two sets of
data is important. 
The \ha\ spectra were calibrated using synoptic observations of a
portion of the North America Nebula (NAN), which has an \ha\ surface
brightness  of \iha = 800 R \citep{Scherb81, Haffner+03}.   
The \hb\ spectra were calibrated by separate \ha\ and \hb\
observations of part of the large \hii\ region surrounding Spica
($\alpha$ Vir), a nearby B1 III star. In the absence of extinction,
the photon number ratio of \iha\ to \ihb\ from  photoionized gas at a
temperature of $T = 8000$ K is 3.94, set by the `Case B' recombination
cascade of hydrogen \citep{Osterbrock89, HS87}.  
We assume that the emission from the Spica  \hii\ region suffers no
extinction because of its proximity \citep[$d \approx 80$
pc;][]{Hipparcos}, high Galactic latitude $b \approx +50\dg$, and the
low interstellar hydrogen column density, $1.0 \times 10^{19} $
cm$^{-2}$, to the exciting star  \citep{YR76}. 
Combining observations of the Spica \hii\ region with NAN, we obtained
a \iha/\ihb\ photon ratio of 5.1 toward NAN, consistent with
observations of extinction toward stars in the nebula
\citep{Cambresy+02}.  
All of the \hb\ spectra were then calibrated assuming an \ihb =
800/5.1 R = 157 R for  NAN. In addition, multiple observations of
standard calibration directions were observed at different zenith
distances  to estimate the nightly transmittance of the atmosphere,
allowing an airmass correction to be applied to data.  

In order to compare accurately the strengths of \nii, \sii, and \oiii\
with \ha\, we needed to apply an additional correction for the difference in transmission of the WHAM instrument at the wavelengths of
these lines. This was done by using the \ha\ and \hb\ calibration, and
assuming that the change in instrumental transmission is linear with
wavelength. This correction is largest for the \oiii\ data (36\%), and very small for \sii\ and \nii\ (3\% and $<$ 1\%, respectively). 
The systematic uncertainty in the response of the
spectrometer at the different wavelengths was not formally
calculated, and may not vary linearly with wavelength. 
However, this level of this uncertainty is below the precision of the physical conditions inferred by the data and discussed in \S\ref{sec:ions}.
In addition, changes in the {\it{ratios}} of these lines, which reflect trends in the physical conditions, are not significantly affected by this source of uncertainty.

\subsection{\ha\ Data}
\label{sec:haobs}

The geocoronal \ha\ emission line is the primary contaminant of the
\ha\ data, with an intensity of $\iha \approx 5-10$ R that varies with
the height of the Earth's solar shadow along the line of sight
\citep{Nossal+01}.  The line has a narrow width of $5-7\ \kms$ (FWHM),
significantly narrower than the 12 \kms\ spectral resolution of WHAM,
and it is at rest with respect to the geocentric reference frame. The
wavelength of this photo-excited transition is slightly different from
that of the \ha\ produced by recombination \citep{Nossal+01}. For
these observations, the geocoronal \ha\ line was significantly
brighter than the relatively wide ($\approx25~\kms$) Galactic emission
features, facilitating the use of the geocoronal line as a velocity
calibration in each spectrum. However, the motion of the Earth with
respect to the local standard of rest (LSR), for the directions and
times of these observations, placed the geocoronal line at a velocity
\vlsr\ between +20 \kms\ and  +40 \kms, obscuring the Galactic
emission at these velocities.  To remove this geocoronal emission from
each spectrum, we modeled it as a single, narrow Gaussian.  A first
estimate of the strength of the line was provided by visual inspection
of the spectrum and theoretical models of the geocoronal line
intensity (S. Nossal, private communication). 

Several fainter atmospheric lines were also present in every
spectrum. These lines have been well characterized through
observations near the Lockman window, the direction with the faintest
Galactic \ha\ and \hi\ emission in the sky \citep{LJM86,
  Hausen+02-apj}. These lines are fixed in a geocentric velocity
frame, have narrow widths of $\approx 5~\kms$,  and typical
intensities of $I  < 0.2$ R. The intensity of the lines varies
proportionally with the airmass of the observation, and their relative
strengths are observed to remain fixed.  To correct each spectrum for
this effect, we followed a procedure similar to that for the WHAM-NSS
reduction.  We used a single atmospheric line template based on data
from \citet{Haffner+03}, with one number, a scaling factor,
parameterizing the underlying atmospheric line spectrum. 

The remaining artifact to be removed from the spectra is the
continuum. 
As mentioned above, some of the spectra have emission that
extends almost all the way across the entire 200 \kms\ window.  As a
result,  estimating the location of the continuum was difficult, and
this significantly contributes to the overall uncertainty of the
resulting interstellar emission profiles. We assume that the continuum is linear in shape \citet{Haffner+03}, and took advantage of the
fact that most of the spectra suffer from relatively high extinction,
and do not appear to contain high-velocity Galactic emission from the more
distant gas. 
These spectral regions, which spanned $\gtrsim$\ 50 \kms, were used to estimate the zero point and slope of the continuum, and were consistent with the range of values found for the WHAM-NSS data \citet{Haffner+03}. 
To estimate the continuum for the spectra that {\it{do}} exhibit high-velocity emission, an average value of the slope and continuum for spectra within the same block, or nearby block observed on the same night, was used.
However, the continuum level and slope is set by the atmosphere as well as the amount of starlight within the beam, which varies from one direction to
another. 

To remove all of these features from the data, we took additional
advantage of the fact that each direction had already been observed
once before in \ha\ during the completion of the WHAM-NSS.  These
survey spectra have been intensity calibrated and corrected for
atmospheric and continuum emission in a more systematic and complete manner, providing an excellent reference for the newer data.  
The Survey data were also taken at a time when the Earth's motion through the LSR placed the geocoronal line at a different velocity in the Galactic spectrum compared to the newer data. 
We simultaneously fitted a geocoronal line, continuum, atmospheric
line template, and wide (FWHM $= 25~\kms$) Galactic emission
components to each spectrum.   
The parameters of the fit were first estimated as discussed above.
The best fit continuum and atmospheric emission lines were then
subtracted, and the resultant pure Galactic spectrum was compared to
the corresponding Survey spectrum at the overlapping velocities
($-50~\kms \lesssim \vlsr \lesssim +100~\kms$).  Small adjustments
(generally less than 10\%) were then made to the strength of the
geocoronal line, the scaling factor for the weaker atmospheric lines,
and the shape of the continuum  to match the shape of the
corresponding Survey spectrum.  The remaining difference between the
newer data and the Survey spectrum, although small ($< 10\%$), form
the largest part of the uncertainty in the \ha\ data. This systematic
source of error comes from small changes in the sensitivity and
flat-field of the instrument over a period of $\sim 4$ years between
when the Survey data and the newer data were taken. 

\subsection{\hb\ Data}

The \hb\ data were corrected in a similar manner as the \ha.  
Each spectrum was modeled as the sum of a linear continuum, a narrow
geocoronal \hb\ line, weak atmospheric lines, and Galactic emission.  
However, there were no pre-existing, calibrated \hb\ data with which
to compare these data.  
As a result, we were unable to remove the atmospheric contaminants and
continuum with same accuracy as the \ha\ data. 

The geocoronal \hb\ line is about 10 times fainter than the geocoronal
\ha\ line \citep{MierkPhD}.   
With $\ihb \approx 0.5 - 1.0$ R, it is about as strong as the Galactic
\hb\ emission components, making its precise position more difficult
to determine and thus complicating the removal of the line.  
A first estimate of the strength and position of the geocoronal \hb\
line was made by a least-squares free fit to the spectra.   
A template of the weaker atmospheric lines was constructed from \hb\
observations near the Lockman window, as was done for the \ha\ data.
A single scaling factor was then used to characterize the strength and
position of these atmospheric lines in the spectra.  
Identifying the continuum proved to be the most uncertain aspect of
the \hb\ data reduction.  
Lacking any pre-existing \hb\ observations, we adopted three values of
the continuum that corresponded to a) the best fit to each spectrum
and b) upper and lower limits as suggested by the data itself. 
The best fit was one that minimized $\chi^2$, and the upper and lower limits placed the data points in the flat portions of the spectra entirely below and above the continuum, respectively. 
This conservative estimate for the continuum uncertainty was incorporated into the overall error estimates, and represents the largest uncertainty in the results.   

A model fit to each spectrum was then used to remove the continuum and
atmospheric emission lines, as was done for the \ha\ data. 
Initial estimates of the parameters of the fit were further
constrained by examining the ratio of the \ha\ to \hb\ spectra.  If
the gas is photoionized, then this ratio should remain constant or
increase with velocity or path length, corresponding to a constant or
increasing amount of dust along the line of sight. 
Small changes ($\approx$\ 10\%) were made to the strength of the geocoronal and
atmospheric lines to ensure that the ratio did not decrease, within the uncertainties, over small velocity intervals that corresponded to the position and widths of the atmospheric lines.  The small changes were within the generally observed fluctuations in atmospheric line intensity over the timescale
of the observations.

\subsection{\nii, \sii, and \oiii\ Data}

Thirteen sightlines were observed in the additional lines of
\nii$~\lambda6583$, \sii$~\lambda6716$, and \oiii$~\lambda5007$. These
sightlines were chosen based on a preliminary examination of the \ha\
data that showed the strongest evidence for bright \ha\ emission at
high velocity. The relative strengths of these lines with respect to
\ha\ can be used to estimate the physical conditions of the emitting
gas, and to search for potential changes in physical conditions with
position in the Galaxy.  The quality of these data are not as high as
the \ha\ and \hb\ data. However, they allow us to search for important
trends in the line ratios and demonstrate the feasibility of a future,
more extensive observational campaign to explore physical conditions
in the inner Galaxy through nebular emission line diagnostic
techniques. 
 
The \nii\ and \sii\ spectra were reduced in a similar manner as the
\ha\ and \hb. They were all taken on one night, with each sightline
observed once for 180 sec. Each observation covered a velocity
interval between -50 \kms\ and +150 \kms\ LSR. To remove the
atmospheric lines from these spectra, a template was used as discussed
in \S\ref{sec:haobs}.  Emission was present across almost the entire
spectrum, and therefore the location of the baseline was difficult to
determine. The strength of the Galactic emission  was always
significantly stronger than the $\lesssim$ 0.1 R atmospheric lines;
therefore, while  estimating the strength of the atmospheric lines was
difficult, the error resulting from this uncertainty was small. To
determine these parameters, a least-squares fit to the spectra was
performed, where the strength of the template and location of the
baseline were changed to minimize the $\chi^2$ of the fit.  The
baseline and template was then subtracted from the spectra. Since the
atmospheric lines were much fainter than the Galactic emission, the
resulting interstellar emission line profiles are not very sensitive
to the particular choice of template strength.  

The \oiii\ data were taken at two different tunes (i.e., two different
positions for the center of the spectrometer's 200~\kms\ passband) to
extend the velocity coverage beyond 200~\kms\ and to help in
characterizing the baseline. Each direction was observed twice at each
tune for 120 sec at a time. The spectra for each sightline at a given
tune were averaged together, with the spectra at the different tunes
stitched together. The velocity calibration was determined using a
semi-empirical prediction based on the pressures of the two etalons.
An attempt was made to remove the atmospheric lines based on a
template. However, for several spectra the Galactic emission intensity
was only comparable to or less than the atmospheric lines, so that
small variations in the unknown strength of the atmospheric lines
yielded very large uncertainties in the Galactic emission. Therefore,
for the \oiii\ spectra,  we adopted an ON-OFF technique, whereby we
subtracted one of the \oiii\ spectra from the rest of the spectra. 
The OFF spectrum that was subtracted from the other
spectra was toward (28\dg, -2.6\dg). This direction is close to the
Galactic plane where the \ha\ and \hb\ data suggested the \av\ is
high, and therefore the Galactic \oiii\ emission is likely very weak,
particularly from the Sagittarius and Scutum arms at \vlsr\ $\gtrsim$
50~\kms. Since this spectrum was taken close to the other spectra in
space and time, the atmospheric lines were accurately removed by this
subtraction. This technique underestimates the Galactic \oiii\
emission, particularly near the LSR, since there may be some Galactic
emission toward the OFF direction.

\section{DIFFUSE, OPTICAL EMISSION FROM THE INNER GALAXY}
\label{sec:emission}

\begin{figure}[htp]
\includegraphics[scale=1.0]{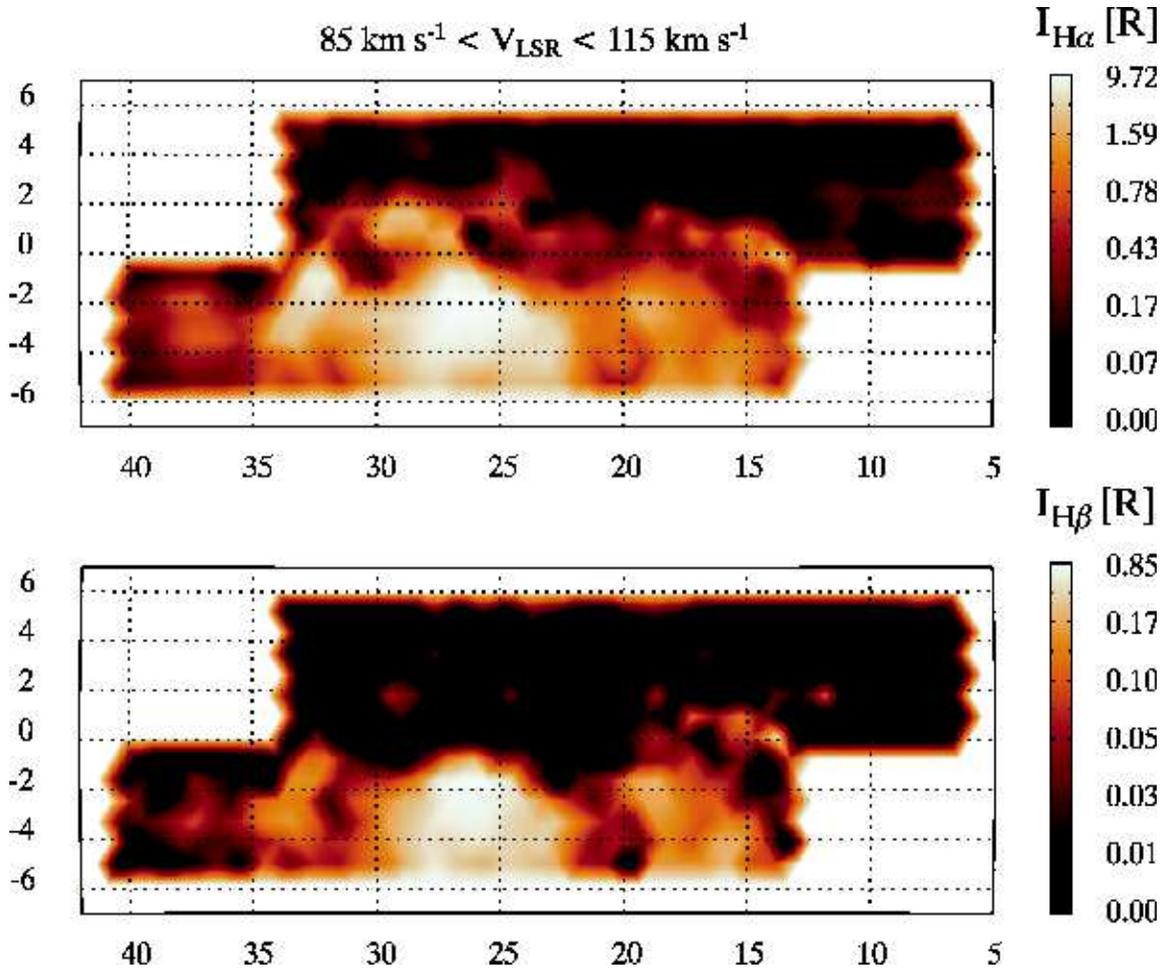}
\caption{ Histogram equalized map of \ha\ and \hb\ emission toward the
  inner Galaxy at high velocity, with +85~\kms~$< \vlsr <$~+115~\kms.
  The spectra near \lb\ = $(27\dg, -3\dg)$ have an emission line
  centered near $\vlsr \approx +100\ \kms$, suggesting that we are
  seeing emission out to the tangent point at a heliocentric distance
  $D_\odot$ $\gtrsim$ 6 kpc.\label{fig:2}} 
\end{figure}

\begin{figure}[htp]
\includegraphics[scale=0.7]{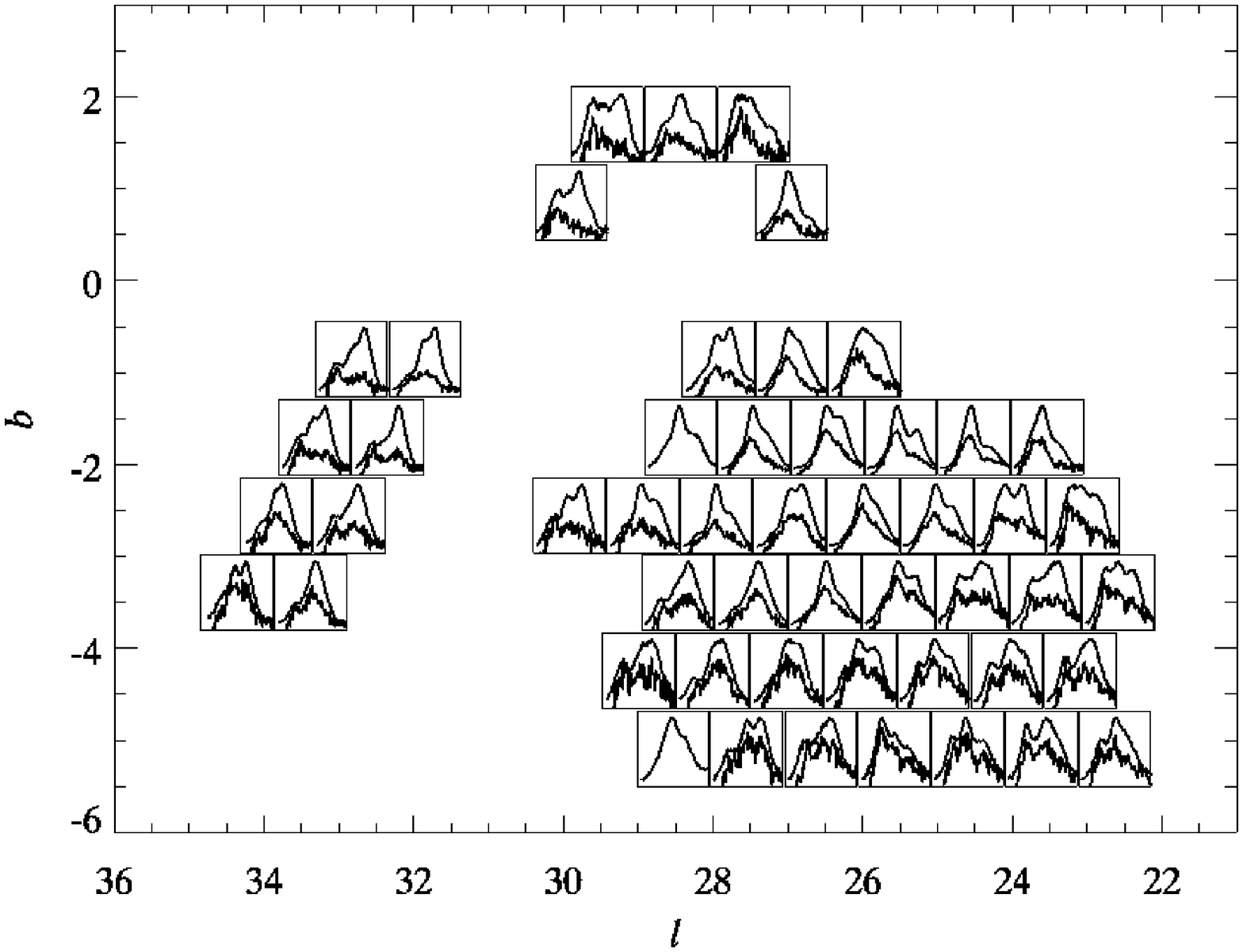}
\caption{ Combination of spectral and spatial information for 51
  directions which show \ha\ emission at \vlsr\ $\gtrsim$ +100 \kms.
  Each square shows a full \ha\ and \hb\ spectrum and location of that
  spectrum on the sky. The emission is shown over a fixed velocity
  range, -50\kms\ to +150\kms\ in each spectrum, with an arbitrary
  vertical scale. The \hb\ spectra have been multiplied by 3.94, the
  expected ratio of \ha/\hb\ in the absence of extinction.  Note the
  increase in \hb/\ha, or decrease in \av, with increasing angular
  distance away from the Galactic plane and at lower \vlsr. For two
  directions no reliable \hb\ spectra were obtained. \label{fig:3}} 
\end{figure}

The spatial distribution of the highest velocity \ha\ and \hb\
emission is shown in Figure~\ref{fig:2}. These velocity channel maps
cover a smaller area than shown in Figure~\ref{fig:1} and highlight
the region surrounding the low extinction window.  Both of the maps
display emission integrated over the velocity range of $+85\ \kms <
\vlsr < +115\ \kms$, revealing a bright region centered near $\lb =
(27\dg, -3\dg)$. The color scaling in Figure~\ref{fig:2} has been
histogram equalized to enhance the contrast of the features in the
map. The morphology of the \ha\ and \hb\ maps are very similar, with a
faint halo of emission extending to the south of the brightest knot,
and a patchy filament of emission near $l=33\dg$. 
The spectra toward this bright region, at longitudes $l \gtrsim
22\dg$, reveal distinct emission components centered near $\vlsr
\approx +80\ \kms$ (see Figure~\ref{fig:3}). The patchy spots of high
velocity emission at lower longitudes ($15\dg < l < 20\dg$), closer to
the Galactic center, are from the wings of much stronger emission
centered at lower velocities. 
At these high velocities, the \hb\ emission relative to the \ha\
emission is about 2.5 times fainter  than that predicted for ionized
gas with no extinction \citep{Osterbrock89, HS87}, indicating a higher
extinction for this gas compared to gas at lower velocities. (see
\S\ref{sec:ext} below).

A combination of spatial and spectral information about this high
velocity emission is shown in Figure~\ref{fig:3}.   Fifty one
directions were selected as having significant emission beyond \vlsr =
+85 \kms, based on a visual inspection of Figure~\ref{fig:2}.  The
\ha\ and \hb\ spectra toward each of these 51 directions is shown,
with their position within this figure given by their Galactic
coordinates.   
The horizontal axis of each spectrum is the radial velocity \vlsr,
spanning a range of $-50\ \kms < \vlsr < +150\ \kms$\ for each
spectrum.  The vertical axes are in arbitrary units, and determined
individually from the intrinsic brightness of the lines.  In two
cases, no reliable \hb\ spectra were available and are omitted from
the figure. The \hb\ spectra are represented by the weaker, noisier
dark lines, and have been multiplied by 3.94, the expected \iha/\ihb\
photon ratio in the absence of extinction.    
Most spectra have two strong, discrete emission features, centered
near $\vlsr \approx +50\ \kms $ and $\vlsr \approx +80\ \kms$.  The
\ha\ and \hb\ spectra decrease in intensity away from the Galactic
plane, and the relative strengths of the \ha\ and \hb\ lines also show
that there is less extinction farther from the plane and at lower
velocities.

\begin{figure}[htp]
\includegraphics[scale=0.7]{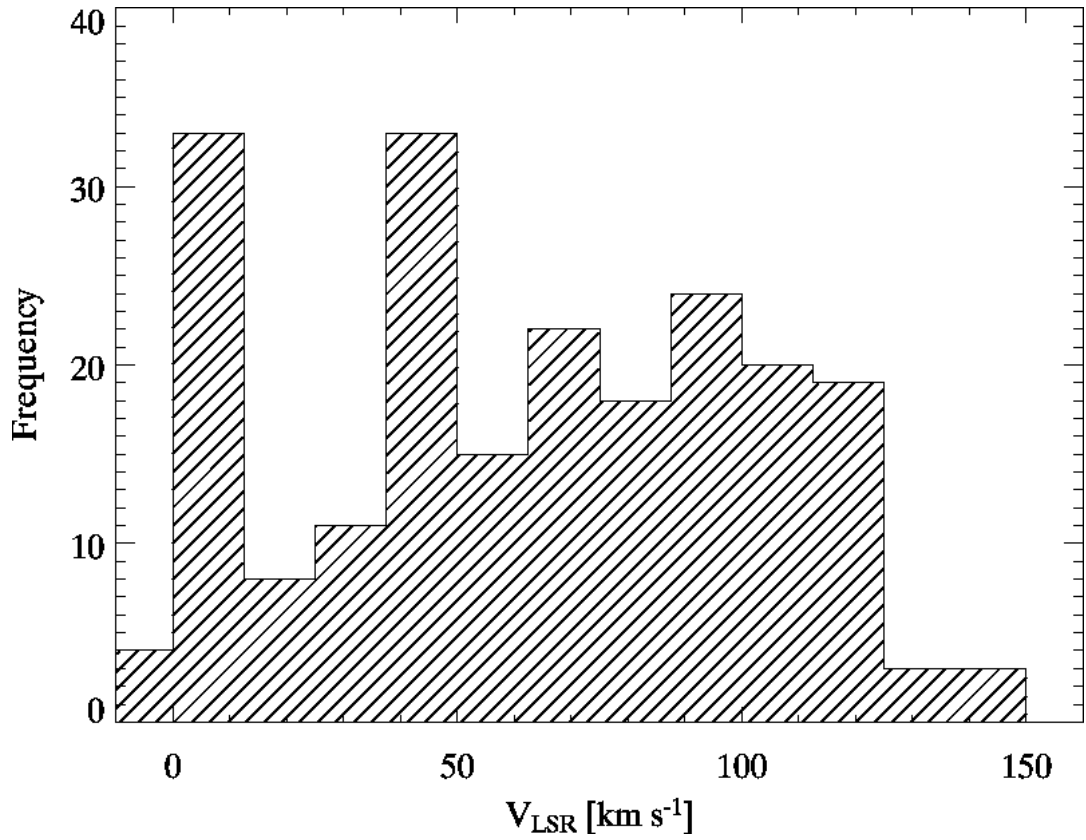}
\caption{ Histogram of the component velocities for the \ha\ spectra
  in the low-extinction window shown in Figure~\ref{fig:3}. The
  velocities were determined through a least-squares fit of Gaussian
  profiles to the spectra.  The width of each bin is 12.5~\kms. There
  are two peaks near 0~\kms\ and +50~\kms, corresponding to emission
  from the solar neighborhood and the Sagittarius spiral arm,
  respectively.  \label{fig:4}} 
\end{figure}

A formal multi-component fit was also made to each of the \ha\ spectra
in Figure~\ref{fig:3}, from which additional information about the
nature of the emission and extinction within the window was obtained
(see \S\ref{sec:ext} below).  Each reduced spectrum was modeled as the
sum of multiple Gaussian components, using a $\chi^2$ minimization
scheme similar to the data reduction procedure in \S\ref{sec:obs}.  
The fitted velocity of each component was recorded, and a histogram of
those velocities appears in Figure~\ref{fig:4}.  On average, five
components were required for a good fit to each spectrum. The
full-width half-max (FWHM) of each component was fixed at $25\ \kms$.
This is the typical width of an \ha\ line in the WHAM-NSS,  and is a
combination of thermal ($\approx 20\ \kms\ {\rm{at}}\ T_e = 10^4~$K)
and non-thermal ($\approx 15\ \kms$) broadening \citep{Haffner+03}. 
The structure of the histogram was not significantly altered when we
allowed the widths of the components to be a free parameter, but not
fixing the widths did complicate the fitting procedure, especially for
faint, overlapping components. The width of the bins in
Figure~\ref{fig:4} is $12.5\ \kms$, half of the FWHM of the components
and also the approximate spectral resolution of the WHAM instrument.

The histogram in Figure~\ref{fig:4} reveals a strong peak near \vlsr\
= 0 \kms, which is emission from the local solar neighborhood.  It is
interesting to note that the peak is slightly red-shifted, consistent
with the gas being just interior to the solar circle at these
longitudes following a flat rotation curve.  There is a second peak
near $\vlsr = +45\ \kms$, which can be interpreted as emission from
the Sagittarius spiral arm.  Beyond $\vlsr \approx +50\ \kms$, there
is a wide range of velocities all the way out to the tangent point
velocity ($\approx +110\ \kms$).  There is marginal evidence for a
concentration of velocity peaks near $\vlsr = +70\ \kms$, the expected
kinematic signature of the Scutum spiral arm. The small number of
velocity peaks at \vlsr~$\ge$~+120\kms\ beyond the tangent point
velocity are associated with very faint ($\lesssim 0.1$ R) emission
and were necessary for a good fit to the data. 
We note, however, these components may not be physical, and could be
part of a higher velocity tail of emission from components at lower
velocities with widths greater than $25\ \kms$.  
On the other hand, at least some of these highest velocity components
do appear to be associated with gas near the tangent point but with a
velocity in excess of that associated with circular rotation
\citep[e.g.,][]{Lockman02b}. 

\section{MEASURING EXTINCTION}
\label{sec:ext}

\subsection{Formalism}

To characterize further the nature of this high-velocity emission
region, we combined the \ha\ and \hb\ data to examine the role of dust
extinction along each line of sight.  
An optically thin photoionized gas at a temperature $T_e=8000$\ K has
an intrinsic line ratio of \iha/\ihb\ = 3.94 in photon units
\citep{Osterbrock89, HS87}; this ratio is only weakly dependent on
temperature ($\propto T^{0.07}$) and density.   
Hydrogen gas ionized by other processes, such as shocks, will not emit
\ha\ and \hb\ photons in the ratio given above. However, strong shocks
from rare supernovae and weaker shocks from spiral arms do not provide
enough power to sustain the ionization observed in the general WIM
\citep{Reynolds90-apj}, and therefore we will assume that the diffuse
gas observed here is photoionized, and that  the recombination ratio
\iha/\ihb\ can be used to infer the extinction. 

Consider a line of sight with an optical depth $\tau_{H\alpha}$ at
\ha\ and $\tau_{H\beta}$ at \hb\ out to a point source of  emission
with an intrinsic brightness $I_{\rm{H}\alpha,int}$ at \ha\ and
$I_{\rm{H}\beta,int}$ at \hb.  The ratio of the intensities of the two
observed lines is 
\begin{equation}
\label{eq:1}
\frac{\iha}{\ihb} = \frac{I_{\rm{H}\alpha, int}\
  e^{-\tau_{H\alpha}}}{I_{\rm{H}\beta,int}\ e^{-\tau_{H\beta}}} 
\end{equation}
We replace the intrinsic line ratio
$I_{\rm{H}\alpha,int}$/$I_{\rm{H}\beta,int}$ with 3.94 from above, to
obtain 
\begin{eqnarray}
\label{eq:2-3}
\mbox{ln}(\frac{\iha/\ihb}{3.94}) & = & \tau_{H\beta} - \tau_{H\alpha} \\
                                      & = & \tau_{H\beta}\, (1 - \frac{\tau_{H\alpha}}{\tau_{H\beta}}).
\end{eqnarray}
The ratio of $\tau_{H\alpha}/\tau_{H\beta}$ depends on the detailed
absorption and scattering characteristics of the dust along the line
of sight. However,  extinction curves measured in the diffuse ISM of
the Galaxy suggest that a fairly well defined relationship exists
between the ratio of extinction at one wavelength compared to
another. \citet{CCM89}, for example, have developed semi-analytical
fits to observed extinction curves which parameterizes these ratios
through a single scaling factor, the total-to-selective extinction
ratio $R_V = A(V) / E(B-V)$.  
For an $R_V = 3.1$, typical of the diffuse ISM, the ratio of the
optical depths $\tau_{H\alpha}/\tau_{H\beta}$ = 0.70.  To convert the
optical depth at \hb\ to the more standard quantity of extinction in
the Johnson $V$ band, we note that the ratio of optical depths
$\tau_{H\beta}/\tau_{V}$ = 1.16. 
With the above relationships, and converting $\tau_V$ into magnitudes
of extinction $A(V)$ through the relation $A(V) = 1.086\, \tau_{V}$,
we obtain $A(V)$ as a function of the observed \iha/\ihb\ line ratio: 
\begin{equation}
\label{eq:4}
A(V) = 3.12\,\mbox{ln}(\frac{\iha/\ihb}{3.94}).
\end{equation}

Note that observations of \iha\ and \ihb\ are equivalent to the color
excess, or the  relative extinction, of the emitting gas.  Using
equation~(\ref{eq:2-3}) with the definition of color excess, we see
that 
ln$[ (\iha/\ihb)  /  3.94] = (1/1.086)\ E(H\beta - H\alpha)$.
Normalized extinction curves in the Galaxy, which quantify the
relationship between $A(\lambda)$ and $1/\lambda$, are linear at
optical wavelengths, and the slope of this line in the optical is not
very sensitive to $R_V$ \citep[see review by][]{Fitzpatrick04}.
Therefore, since the difference in the inverse wavelengths between
\ha\ and \hb, is about the same as the difference between the Johnson
$B$ and $V$ bands, the color excess $E(H\beta - H\alpha)$ is similar
in value to the commonly measured color excess $E(B-V)$. Using the
analytic formalism of \citet{CCM89}, the ratio of the two color
excesses are 
\begin{equation}
\label{eq:5}
\frac{E(H\alpha - H\beta)}{E(B-V)} = 0.12R_V + 0.73
\end{equation}
Thus $E(H\alpha - H\beta)$ and $E(B-V)$ differ only by 10\% for an
$R_V = 3.1$, and never depart from each other by more than 30\% for
all values of $R_V$ in the Galaxy ($2 < R_V < 5$). 
As a result, observations of \iha/\ihb\ can be used to infer the
equivalent value of the color excess \ebv\ with an uncertainty of
$\approx 10\%$. 
However, a value of $R_V$ is required in order to quantify the
\emph{total} extinction to the emitting source, $A(V)$. 

We note that there is growing evidence that $R_V$ may be systematically lower in the inner Galaxy \citep{Popowski00, Udalski03}.  \ha, \hb, and radio observations of planetary nebulae toward the Galactic bulge suggest that $R_V$ may be as low as 2.0 \citep{WBC93, Ruffle+04}.  However, there is considerable uncertainty in the distances to these PNe, their association with the bulge, and the derived values of $R_V$.  Our discussion of \av\ below assumes a standard $R_V$ = 3.1, but we note that this value may decrease with Galactocentric radius.

\subsection{Caveats}

The discussion above holds true in cases where the \ha\ and \hb\
emission comes from a point source and is subsequently removed from
the observer's beam through dust absorption and scattering out of the
line of sight. Our observations, however, are of \emph{diffuse}
emission. In this case, dust can scatter diffuse light originating
from another direction back \emph{into} our 1\dg\ diameter beam and
\iha/\ihb\ is a measure of the attenuation rather than
extinction. This scattering effect complicates the conversion between
the observed \iha/\ihb\ line ratio and the standard definition of
interstellar extinction.  In order to quantify this effect accurately,
detailed radiative transfer models are required, which incorporate the
(unknown) distribution of dust and emission, as well as
poorly-constrained dust scattering properties
\citep[e.g.,][]{Mathis83, Gordon04}. 

However, this effect is expected to be minimal for emission sources
that subtend solid angles that are small compared to the typical
scattering angle, which is $\sim 50\dg$ in the optical
\citep{Gordon04}.  Consider a patch of emission in the sky at some
distance, with diffuse emission isolated to only that patch in the
sky. The solid angle subtended by this emission, as viewed from an
average dust particle between the observer and the source, is small.  
For a patch of emission with an angular diameter of a few degrees, at
a distance of a few kiloparsecs, almost all of the dust along the line
of sight absorbs and scatters the emission as if it were coming from a
point source. 
As a result, the relationship between \av\ and \iha/\ihb\ in
equation~(\ref{eq:4}) is accurate for measurements toward regions in
which the \ha\ and \hb\ emission  region is limited to a small region
of the sky, with the intervening dust distributed along a large
heliocentric distance, as is the case with the high velocity emission
toward the Scutum cloud. 
This accuracy diminishes with decreasing distance between the observer
and the source, such as for the very extended emission from the local
neighborhood near \vlsr\ = 0~\kms.  

There is an additional complication that must be considered in the
relation between \iha/\ihb\ and \av. In equation (\ref{eq:4}), \iha\
and \ihb\ represent the total emission from a source,
obtained by integrating the total number of \ha\ and \hb\ photons
emitted from a single interstellar cloud at some distance, from which
an \av\ to the cloud may be measured.  However, our observations are
of emission along an extended line of sight with sources at different
velocities (distances). Each line of sight intersects multiple parcels
of warm ionized gas, such as spiral arms, and each parcel has  a FWHM
$\approx 25~\kms$.  Isolating which photons are coming from which
`cloud' becomes formidable if the clouds are not sufficiently
separated in velocity.  Therefore care must be taken in interpreting
the value of \av\ at any one velocity point, especially at a velocity
between well-defined peaks in the emission line profile or for
profiles in which no strong peaks can be identified.  In this case,
the \emph{average} value of \av\ over a velocity interval
corresponding to the expected width of the line is more representative
of the actual value of \av.  

\subsection{Verification}

\begin{figure}[htp]
\epsscale{0.75}
\includegraphics[scale=0.75]{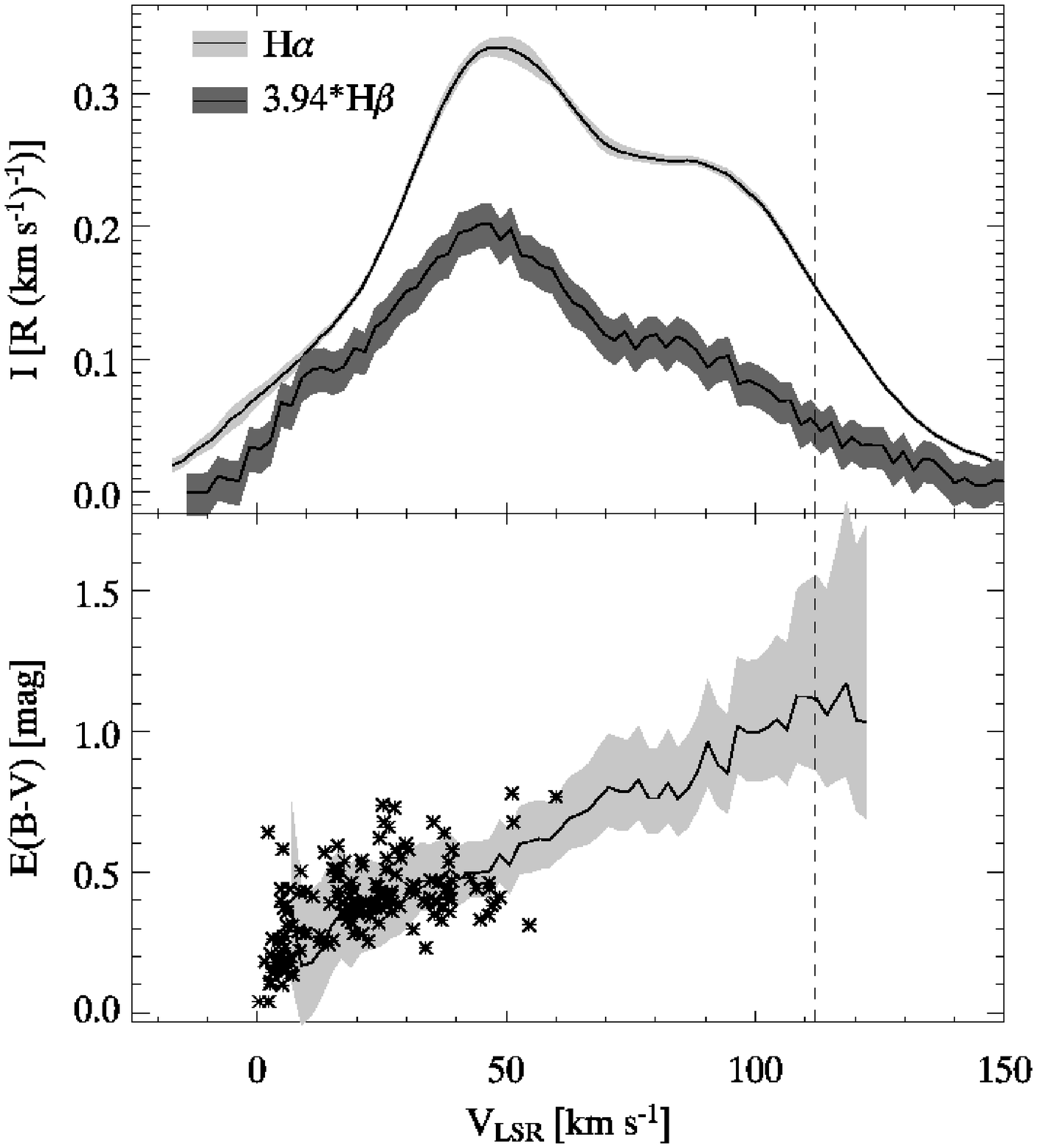}
\caption{ Comparison of extinction measurements with those of
  \citet{Reichen+90} toward the Scutum cloud. The top panel shows the
  average of nine \ha\ and \hb\ spectra within the low \ebv\ area
  outlined by \citep{Reichen+90}, with the shaded regions representing
  the systematic uncertainties. The bottom panel shows the inferred
  value of \ebv\ from the \ha\ and \hb\ spectra. The asterisks are
  measured values of \ebv\ for several UV bright stars in this area,
  where the distance to each star was converted to an LSR velocity
  using the rotation curve of \citet{Clemens85}.  The dashed vertical
  line is the location of the tangent point velocity. We find that the
  two data sets agree within the uncertainties.\label{fig:5}} 
\end{figure}

We can verify the relationship between \iha/\ihb\ and extinction by
comparing our estimates of \ebv\ with direct measurements from stellar
spectrophotometry in the same region of the Galaxy. \citet{Reichen+90}
used a wide-field, balloon-borne UV imager to study a 6\dg\ region
centered on the Scutum cloud. Using a combination of photometric and
spectroscopic distances, they mapped out the changes in \ebv\ with
distance toward stars they detected in the UV.  They identified a low
extinction window within their 6\dg\ field, which is coincident with
the directions in which we see high-velocity \ha\ and \hb\ emission.  
The average of nine WHAM \ha\ and \hb\ spectra that lie in this
window, as defined by \citet{Reichen+90}, is shown in the top panel of
Figure~\ref{fig:5}.

The shaded regions around the \ha\ and \hb\ spectrum in
Figure~\ref{fig:5} represent the maximum extent of the uncertainty in
the data, combining random and systematic uncertainties.  Since the
uncertainty in the \ha\ data comes primarily from systematic
uncertainty in the calibration, the uncertainty in one spectrum is
correlated with the uncertainty in another. Therefore, the uncertainty
in the average \ha\ spectrum shown in Figure~\ref{fig:5} was estimated
by taking the difference between the average of the upper limits and
the average of the lower limits of all of the spectra. 
The uncertainty in the \hb\ spectra, on the other hand, comes
primarily from the uncertainty in the location of the
continuum. Assuming that the uncertainty in the location of this
baseline is uncorrelated from one direction to another, we estimated
the overall uncertainty in the average spectrum to be the average
uncertainty of the continuum placement divided by the square root of
the number of spectra (9) used to construct the \hb\ spectrum in
Figure~\ref{fig:5}. 
This same method is used for of the all of figures in the paper in
which average \ha\ and \hb\ spectra appear. It is important to note
that the shaded regions do not represent Poisson errors, but rather
the maximum allowable values of the data, given the random and
systematic uncertainties. 

The bottom panel of Figure~\ref{fig:5} shows our inferred values of
the color excess \ebv\  as a function of velocity, using $R_V = 3.1$. 
The shaded region around the dark line is the uncertainty in our
inferred value of \ebv, calculated by combining the upper and lower
limits of the average \ha\ and \hb\  spectrum.  Only those radial
velocities with reasonably low values of the uncertainty are shown,
which excludes \vlsr~$\lesssim 0~\kms$\ and \vlsr~$\gtrsim
+120~\kms$. 
The asterisks are from the data for the UV bright stars from
\citet{Reichen+90}, and have been placed on this diagram using a
distance-velocity conversion from the Galactic rotation curve of
\citet{Clemens85} with $R_{\odot} = 8.5$ kpc.  This rotation curve is
used throughout the paper. The vertical dashed line indicates the
location of the tangent point velocity. Photometric error bars and
distance uncertainties for the stellar data are not provided in
\citet{Reichen+90}, but they are likely to be large for the faint,
distant stars at $D \gtrsim 2\ $kpc, corresponding to a \vlsr~$\gtrsim
+40~\kms$.  
We note that the scatter in the stellar \ebv\ data is larger than the
uncertainty in the velocity-distance relation that may be caused by
deviations from non-circular rotation \citep{Lockman02}. Therefore,
the scatter is likely related to either the patchiness of the extinction
within this region or observational uncertainties.
We find that our estimates of \ebv\ match those observed by
\citet{Reichen+90}, within this scatter.  
 
\subsection{Results}
\label{sec:ext_res}

\begin{figure}[htp]
\includegraphics[scale=0.7]{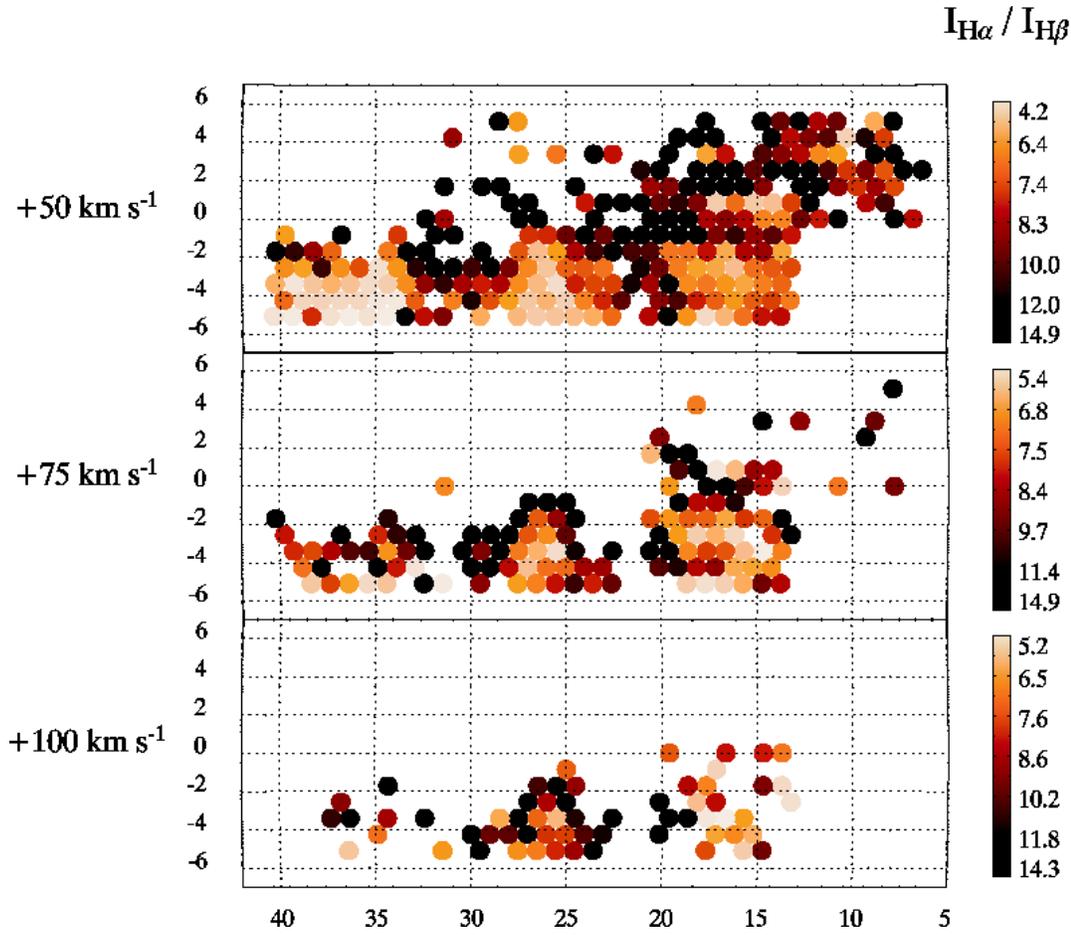}
\caption{ Three velocity channel maps of the average ratio of
  \iha/\ihb\ at \vlsr = +50 \kms, +75 \kms, and +100 \kms, over a
  width of 25 \kms. Observations with large uncertainties have been
  omitted. Note the area of low \iha/\ihb\ near (27\dg, -3\dg), in
  which the extinction remains low out to near the tangent point
  velocity. \label{fig:6}} 
\end{figure}

The assumptions that enter into the conversion between \iha/\ihb\ and
extinction appear valid, and  we can therefore use with confidence our
observations to estimate extinction toward the inner Galaxy, well
beyond that provided by stellar measurements (see Fig.~\ref{fig:5}). 
Figure~\ref{fig:6} illustrates the low extinction in the window toward
the Scutum cloud. 
The three maps are of the average value of \iha/\ihb, calculated by
integrating the emission over a $25\ \kms$ interval, centered at the
three indicated velocities.  
These three velocities are near the location of peaks in the emission
profiles toward the Scutum cloud, and are associated with the
Sagittarius arm, Scutum arm, and the tangent point velocity. Data with
insufficient signal-to-noise have been omitted from the figure.   The
color scaling has been histogram equalized to enhance the contrast in
the maps. 
The top panel, centered at $\vlsr=+50\ \kms$, shows several regions
with relatively low values of \iha/\ihb\ located below the plane near
$l \approx 17\dg, 26\dg, \rm{and}~37\dg$.   
The lower two panels, centered at $\vlsr=+75~\kms$ and +100~\kms,
have much fewer data points due to the lack of measureable \ha\ and
\hb\ emission beyond this velocity in many directions, at least in
part because of significantly increased extinction. The low
extinction region near \lb\ = (26\dg, -3\dg) is apparent, which
persists in the bottom panel centered at $\vlsr = +100~\kms$.

The maps in Figure~\ref{fig:6} can be used to infer the values of \av\
out to different distances, in different locations in the inner
Galaxy. However, for directions away from the Scutum cloud, which is
near $l=26\dg$, the kinematics of the Galaxy displace the centroids of the emission lines, from gas at the same distance, toward different velocities. Some of these velocities are outside the integration range of each panel in Figure~\ref{fig:6}.  
In these directions, the ratio \iha/\ihb\ may be an average over the
\emph{wing} of a line, and the conversion to values of \av\ and its
interpretation, based on the maps in Figure~\ref{fig:6}, become
unclear. 
It is only toward the Scutum Cloud, where emission lines are centered near the velocities in each panel, that \av\ can be reliably inferred from these maps.
The highest velocity components toward this window, centered near \vlsr = +100\kms, place the emission at a kinematic distance $D_\odot \gtrsim 6$~kpc. 
From Figure~\ref{fig:6}, we see the average ratio \iha/\ihb\ centered
at this high velocity is $\approx$\ 10, which corresponds to an $\av
\approx 3$ at this distance (see also Fig.~\ref{fig:5}).

As a further illustration of this low extinction window, sample
spectra toward the Scutum Cloud are shown in Figure~\ref{fig:7}.  The
top panels show the \ha\ and \hb\ spectra, with the shaded regions
representing their respective uncertainties, as in
Figure~\ref{fig:5}. The \ha\ spectra are the high intensity, smoother
profiles, with the axis labeled on the left side of the panel. The
\hb\ spectra are the weaker, noisier profiles, with the axis labeled
to the right of the panel. The next two panels show the average \hi\
and CO emission in these directions, from the surveys of
\citet{HIAtlas} and \citet{DHT01}, respectively.
The angular resolution of the \hi\ and CO surveys are 0.5\dg, and the spectra in Figure~\ref{fig:7} are averages of the HI and CO observations that lie within the larger 1\dg WHAM beam.
The panel below the CO spectrum shows the \ha\ to \hb\ intensity ratio, and the bottom panel shows the inferred values of \av\ at each velocity point from
equation \ref{eq:5}, with the uncertainties indicated by the shaded
regions. Similar to Figure~\ref{fig:5}, these panels only show
\ha/\hb\ and \av\ over velocities in which the uncertainties remain
relatively low (i.e., data beyond the tangent point velocity and at
\vlsr~$<$~0\kms\ are not plotted). The horizontal axes are the
observable quantity \vlsr\ (bottom of panel) as well as the inferred
corresponding kinematic distance (top of panel). The vertical
dot-dashed line running through all the panels indicates the location
of the tangent point velocity.  The horizontal dashed line in the
\ha/\hb\ plots is at 3.94, the expected ratio in the absence of
extinction.  
Note that in all of the spectra, \av\ remains flat or increases with
velocity, as expected from the cumulative effects of attenuation with
distance. In two cases, \av\ starts near zero,  providing additional
confidence in both the data reduction procedure and the conversion
from \iha/\ihb\ to \av.  Note that some directions exhibit a steep
rise in \av, suggesting the presence of a strong concentration of dust
at the corresponding distance (see \S\ref{sec:neutral}). 

\begin{figure}[htp]
\includegraphics[scale=0.7]{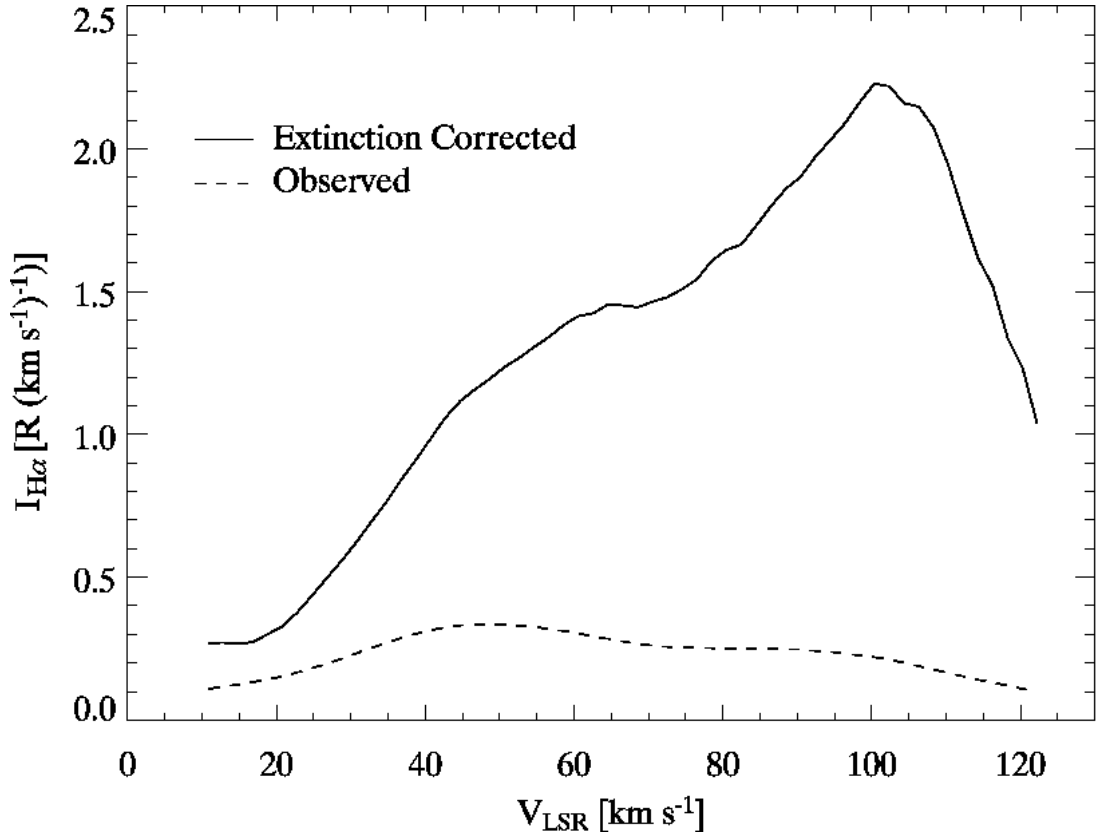}
\caption{Comparison of the observed and extinction corrected \ha\
  emission toward the window. The observed spectrum is the average of
  the same nine spectra used in Figure~\ref{fig:5}. The optical depth
  at each velocity data point was inferred from the average ratio of
  \iha/\ihb\ in a 25\kms\ window around each data point (see
  \S\ref{sec:ext_res}). Note the particularly strong emission at
  \vlsr\ $\approx$\ +105 \kms, which is near the tangent point
  velocity.  \label{fig:hacor}} 
\end{figure}

The measured values of \av\ at each velocity point allows us to
correct our observed \ha\ profiles for
extinction. Figure~\ref{fig:hacor} shows two spectra, one of which has
been corrected for extinction. The `observed' spectrum is an average
of nine spectra toward the window, and is the same \ha\ spectrum from
Figure~\ref{fig:5}. Each observed data point was multiplied by a
correction factor of $e^{\tau_{H\alpha}}$, where $\tau_{H\alpha}$ was
determined from the \hb\ data. In order to produce a smooth profile in
the corrected spectrum, the average value of \iha\ and \ihb\ over a 25
\kms\ interval for each data point was used to compute
$\tau_{H\alpha}$. Figure~\ref{fig:hacor} shows the magnitude of the effect
of this extinction correction, and shows that, on average, there may be
more emission at larger heliocentric distances near the tangent point,
compared to the solar neighborhood and intervening spiral arms. 
Because of the ambiguity in the velocity-distance relationship in the inner Galaxy, some of the emission near the tangent point could be from gas that lies beyond the tangent point. However, this gas is at a larger Galactocentric distance and z-height, where the density, and contribution the corrected emission measure, is likely to be lower. 
The corrected spectrum maintains its multiple component features, implying
enhanced emission from the Sagittarius and Scutum spiral arms and from
the tangent point.  

\section{SCALE HEIGHT AND DENSITY OF IONIZED GAS}
\label{sec:scale}

Correcting the observed \ha\ emission for extinction allows us to
infer some intrinsic properties of the ionized gas in the inner
Galaxy.   
Since the \ha\ emission is related to the emission measure of the gas,
$EM = \int n_e^2 dl$, we can extract information about the density and
the vertical distribution of H$^+$ by  examining the change in the
measured \iha\ in both latitude and distance toward the window. 
Following the formalism of \citet{HRT99}, we assume that the ionized
gas has a vertical distribution about the Galactic plane with the form
\begin{equation}
\label{eq:6}
n_e(z) = n_e^0~e^{-|z|/H}~\rm{cm}^{-3}
\end{equation}
where $n_e^0$ is the density in the midplane and $H$ is the scale height.
The relationship between the extinction-corrrected \ha\ intensity and
the gas density is  
\begin{equation}
\label{eq:7}
2.75 T_4^{0.9} \iha = \int n_e^2 dl, 
\end{equation}
where $T_4$ is the electron temperature of the gas in units of $10^4$
K, and \iha\ is measured in Rayleighs.
If we use $\phi$ to denote the line of sight filling fraction of the emitting gas, and assume that the temperature, filling fraction, and path length
through an ionized region are not functions of vertical
height $z$, then we can rewrite equation~(\ref{eq:7}) as 
\begin{equation}
\label{eq:8}
\iha = \frac{\phi (n_e^0)^2 L}{2.75T_4^{0.9}} e^{-2|z|/H} = I_{\rm{H}\alpha}^0~e^{-2|z|/H},
\end{equation}
where $I_{\rm{H}\alpha}^0$ is the extinction-corrected \ha\ intensity
at the midplane. 
Finally, we replace $z$ with $D_\odot\ \rm{tan}|b|$, where $D_\odot$
is the heliocentric distance to the emitting gas at a Galactic
latitude $b$ to obtain 
\begin{equation}
\label{eq:9}
{\rm{ln}}\ \iha = {\rm{ln}}\ I_{\rm{H}\alpha}^0 - \frac{2D_\odot}{H}{\rm{tan}}\ |b|.
\end{equation}
We now have a relationship between two observable quantities, \iha\
and $b$, from which we may infer the root mean square (rms) midplane
density $\phi^{1/2} n_e^0$ and scale height $H$.  
Additionally, the low extinction window allows us to isolate emission
from different distances and estimate the density and scale height as
a function of Galactic radius. 

\begin{figure}[htp]
\includegraphics[scale=0.7]{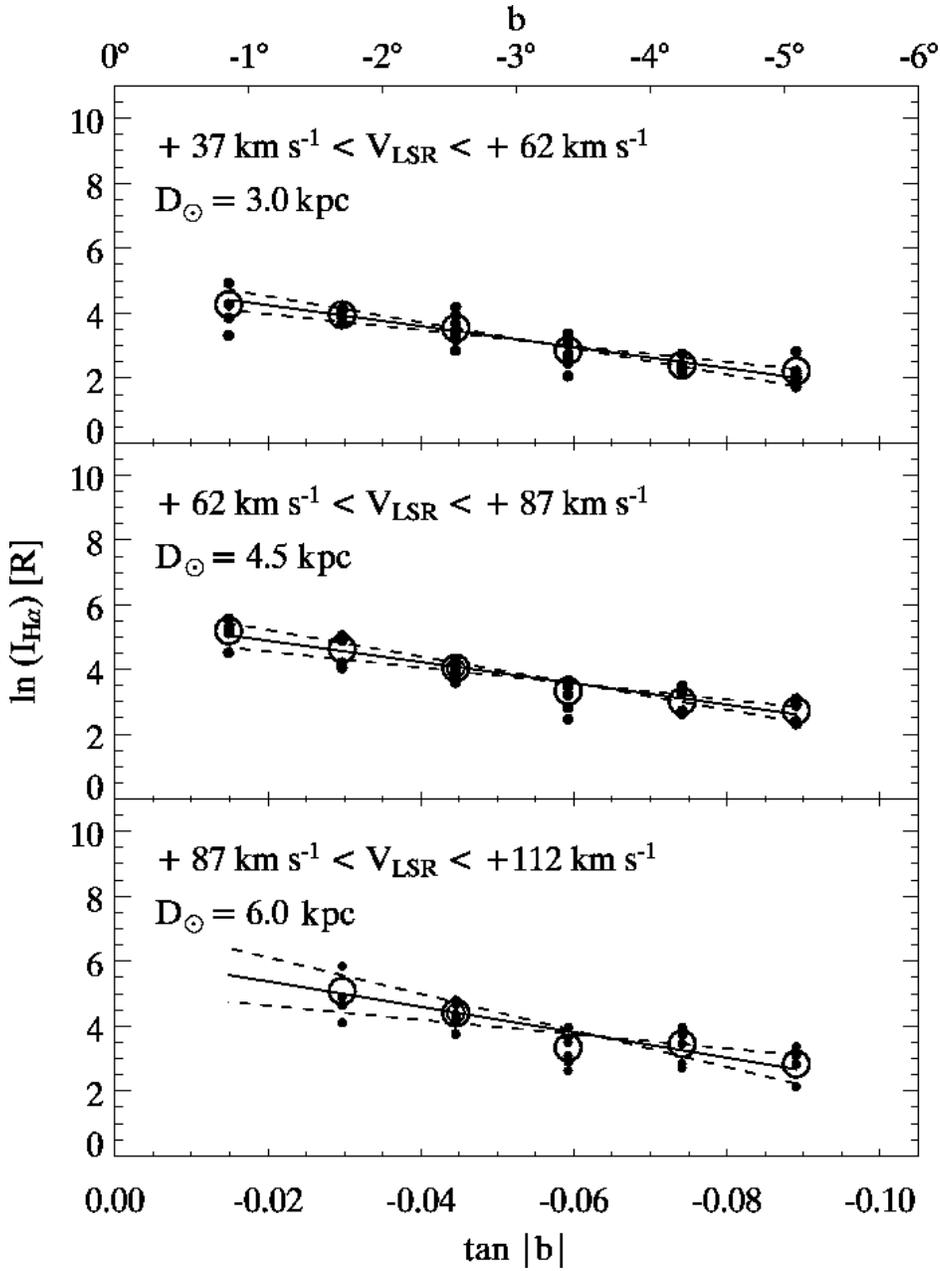}
\caption{ Extinction-corrected \ha\ emission vs. angular distance from
  the Galactic plane for emission at three different radial velocity
  intervals. The assumed heliocentric distances $D_\odot$ to the
  emission regions are also shown. A best fit line to the data and the
  uncertainty is shown as solid and dashed lines, respectively (see
  \S\ref{sec:scale}). \label{fig:8}} 
\end{figure}

\begin{figure}[htp]
\includegraphics[scale=0.7, trim=-1in 0in 0in 0in]{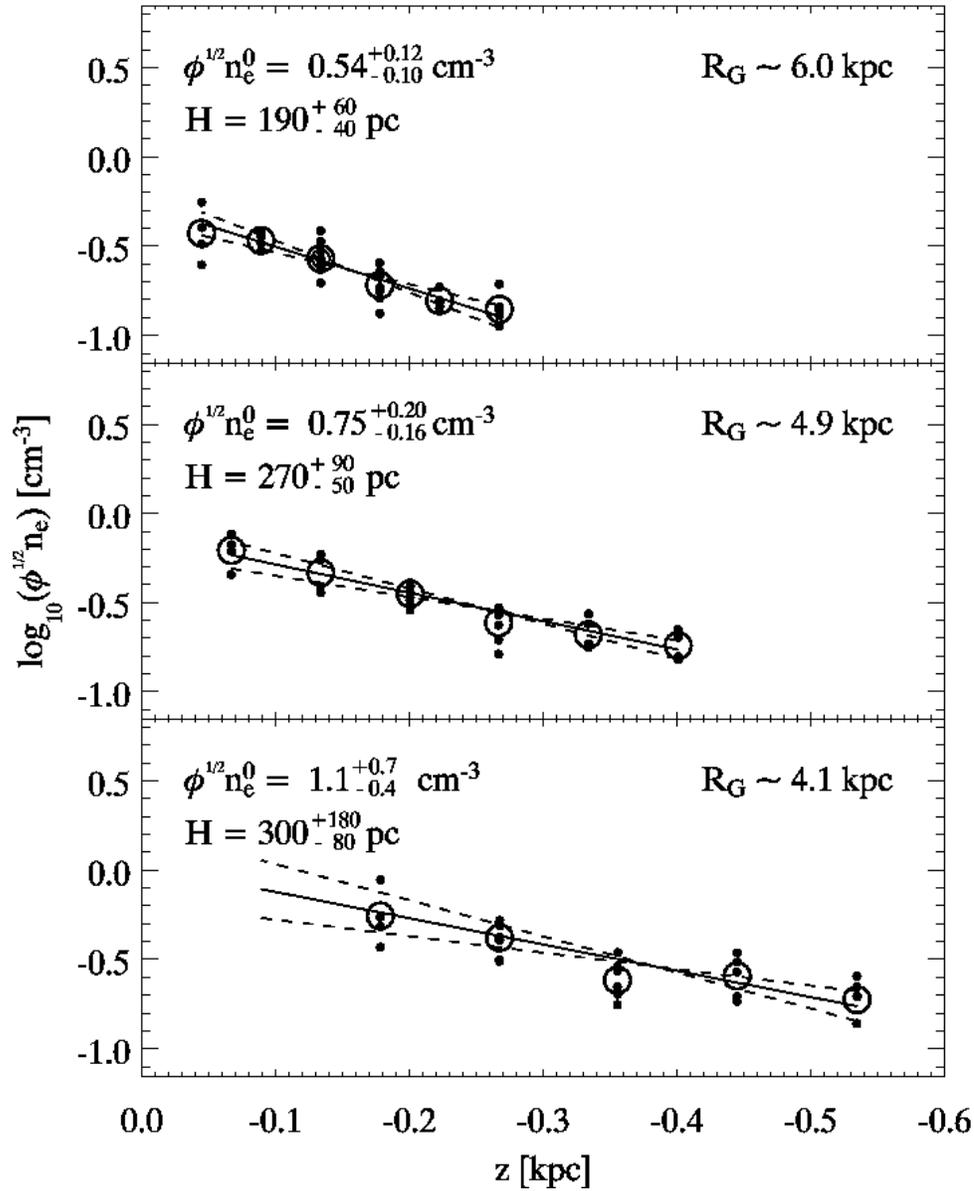}
\caption{ Same as Figure~\ref{fig:8}, except the observable quantities
  of \iha\ and $b$ have been converted into the physical quantities of
  density $n_e$ and vertical height $z$, as described in
  \S\ref{sec:scale}. The best-fit values for the rms midplane density
  $\phi^{1/2} n_e^0$ and scale height $H$ are shown in the upper
  left. The assumed distance from the center of the Galaxy to the gas
  described by these physical parameters is also shown.  
\label{fig:9}} 
\end{figure}

Results of these calculations are shown in Figures~\ref{fig:8} and
\ref{fig:9}. Figure~\ref{fig:8} shows the observed quantities of
(extinction-corrected) \iha\ versus~$b$, while Figure~\ref{fig:9}
illustrates the inferred quantities of $\phi^{1/2} n_e^0$ and $H$
versus $z$-height. 
These data are from spectra toward the low extinction window shown in
Figure~\ref{fig:3} that lie below the Galactic plane.  In order to
perform a reliable extinction correction, only those spectra whose
integrated intensity exceeded our average systematic uncertainty of
0.1 R were considered.  The inferred values of \av\ as a function of
velocity were converted into an optical depth, $\tau_{H\alpha}$, at
each velocity point by taking the average value of \iha/\ihb\ within a
$25\ \kms$ interval around each point and applying the relationship in
equation~(\ref{eq:4}). This averaging was required because of the
finite width of the emission lines, and because some averaging is
needed to produce smooth profiles at low intensities.

\begin{figure}[htp]
\includegraphics*[scale=0.7]{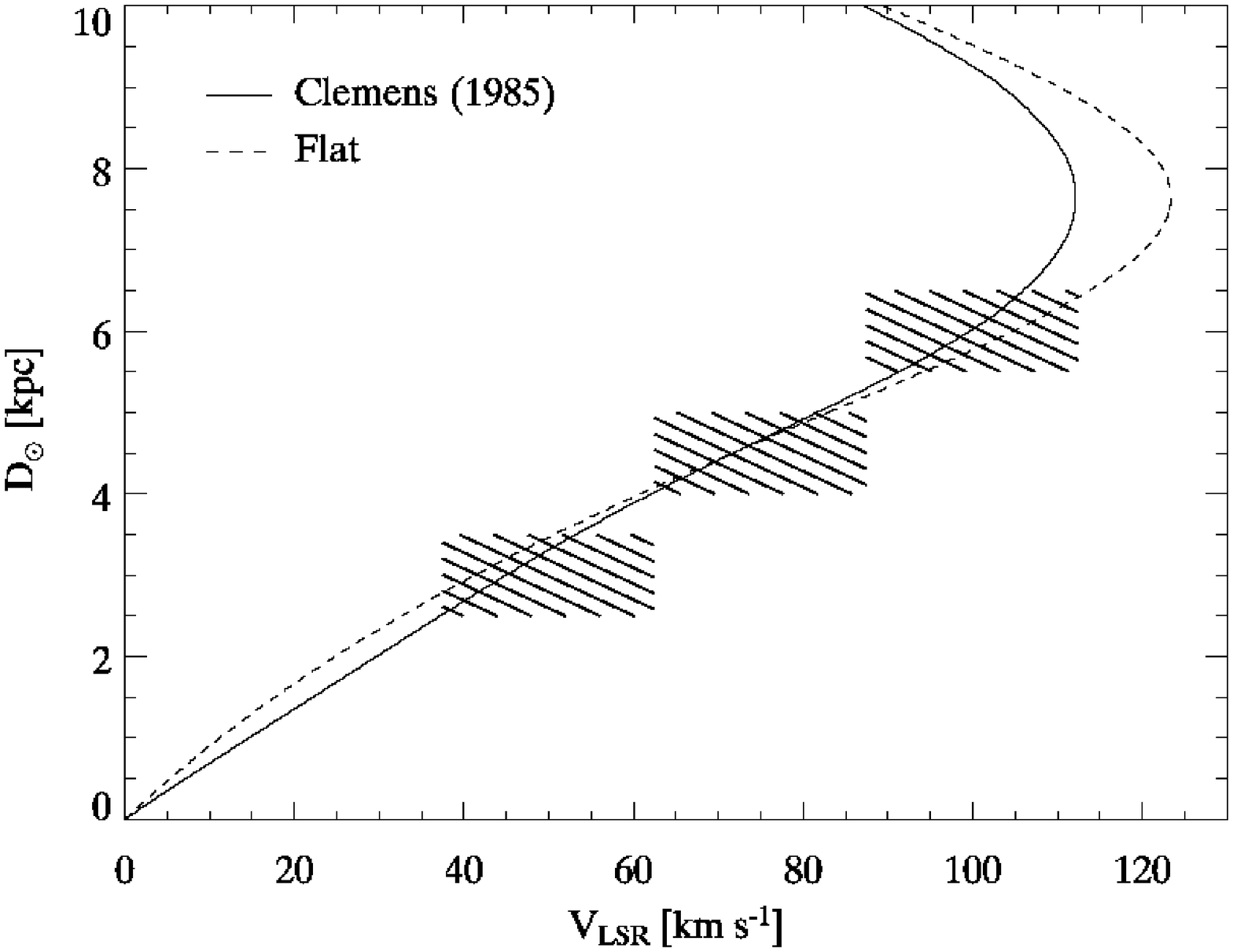}
\caption{Relationship between LSR velocity and heliocentric distance
  toward $\lb = (26.5\dg, -2.5\dg)$ for two different rotation
  curves. The rotation curve from \citet{Clemens85} and a flat curve,
  assuming $\Omega = 220~\kms$ and $R_\odot = 8.5$ kpc, are
  represented by the solid and dashed lines, respectively. The three
  hatched regions show the extent of our assumed distances to the \ha\
  emission over three different velocity intervals, as discussed in
  \S\ref{sec:scale}.  \label{fig:kindist}} 
\end{figure}

In order to convert the observed quantities shown in
Figure~\ref{fig:8} to the inferred quantities in Figure~\ref{fig:9},
the distance to the emitting gas $D_\odot$ and its pathlength $L$ must
be known.  
Each panel in Figure~\ref{fig:8} shows the \ha\ emission integrated
over the three velocity intervals, each 25~\kms\ wide, that are
centered at $\vlsr=+50, +75, \rm{and}\ +100\ \kms$.  These intervals
encompass the location of the fitted velocity components shown in
Figure~\ref{fig:4} and represent emission from the Sagittarius arm,
Scutum arm, and near the tangent point. The emission within each
interval is assumed to occupy a pathlength $L=1$ kpc along the line of
sight, centered at heliocentric distances $D_\odot$ = 3.0, 4.5, and
6.0 kpc, respectively. These distances are consistent with the
standard spiral arm model of \cite{TC93} and the Galactic rotation
curve model from \citet{Clemens85} with $R_{\odot}=8.5$ kpc, as shown
in Figure~\ref{fig:kindist}.  Figure~\ref{fig:kindist} also includes
the relationship between \vlsr\ and $D_\odot$ for a flat rotation
curve, and shows that the kinematic distances we adopt are not very
sensitive to the particular rotation curve that is assumed.   
We see that the assumed path length, $L=1$ kpc, is the approximate
change in kinematic distance over a $25\ \kms$ interval for all three
selected velocity intervals. We note, however, that non-circular
rotation as well as random motions associated with spiral arms,
expanding shells, and shear motion may be present along these lines of
sight. The magnitude of these effects are generally thought to produce
distance uncertainties of less than 10\%, but may vary by more than
this amount for any individual sight line \citep{Lockman02}. A
definitive conversion between emission at a particular velocity and a
specific distance therefore suffers from this uncertainty, and our
adopted distances should be understood as estimates.  From
equation~(\ref{eq:9}) above, we note that the inferred scale height
$H$ is directly proportional to  the adopted distance $D_\odot$.  

In addition, it may be possible that we are detecting emission from
ionized gas beyond the tangent point. This extremely distant emission
would appear at lower velocities with increasing distance and, if
present, would affect the $D_\odot = 6.0$ kpc emission most, where the
velocity `turn-around' occurs (see Figure~\ref{fig:kindist}).  Since
the overall emission decreases with increasing distance from the
plane, the emission from beyond the tangent point, if present, would
result in an overestimate of the emission at lower latitudes.  
However, because the emission beyond the tangent point, at a fixed
latitude, is farther away from the plane, it is expected to be much
fainter than the emission closer to the Sun. We will assume that this
effect is negligible for the $D_\odot=3.0\ \rm{and}\ 4.5$ kpc
emission, and that for the $D_\odot=6.0$ kpc emission, the inferred
midplane density may be overestimated and the scale height may be
underestimated. 

For each latitude in Figure~\ref{fig:8}, there are 2-5 sightlines and
thus a range of values for the extinction corrected \iha. The small,
filled circles in each panel represent individual observations at
different longitudes within the low extinction window.  The larger,
open circles represent the average value of the individual
observations at each latitude. Because there is higher extinction
closer the plane, the extinction correction systematically steepens
the slope of the data points.   
The filled circles provide a sense of the magnitude of the variation in
the data at each latitude, which is of the same order or smaller than
the formal uncertainty of individual observations. 
The solid line in the figure represents a first-order least-squares
fit to the open circles. To estimate the uncertainty in the fit, we
assigned a 1$\sigma$ uncertainty to each filled circle equal to the
maximum deviation of the filled circles.  
The dashed lines in the figure show the extent of the $\pm 1\sigma$
uncertainties in the fitted slope ($-2D_\odot/H$) and intercept (ln
$I_{\rm{H}\alpha}^{0}$). 
 
Figure~\ref{fig:9} shows the values of these best-fit parameters, with
their $1\sigma$ uncertainties, converted into the rms midplane density
$\phi^{1/2} n_e^0$ and scale height $H$.  Here, the vertical axis has
been changed to logarithm of the inferred free electron density, using
equation~(\ref{eq:8}) and assuming $T_e = 8000$ K and $L = 1$ kpc.
The horizontal axis has been replaced with distance from the Galactic
plane, using the adopted distances $D_\odot$ from above.  The distance
to the emission is also expressed with respect to the center of the
Galaxy, $R_G$, using the average longitude of the data, $<l>\ \approx
26\dg$. 
The best fit rms midplane density increases from  0.54 cm$^{-3}$ at
$R_G \sim 6.0\ $kpc to 1.1 cm$^{-3}$ at $R_G \sim 4.1\ $kpc. We note
that the conversion to the actual space density $n_e$ within the
ionized gas suffers from uncertainty in the poorly constrained path
length and filling factor, with $n_e \propto (\phi~L)^{-1/2}$.   
The value of $\phi$ is presently unconstrained. 
Observations in the neighborhood of the Sun of the warm ionized
medium, which has a significantly lower rms density (0.05 cm$^{-3}$)
and larger scale height (1 kpc) than the ionized gas found here, show
that it has a filling factor of $\phi \approx 0.1 - 0.3$
\citep{Reynolds91a, NCT92, MBM04}.   

If the product $\phi L$ does not change considerably with distance,
then the data in Figure~\ref{fig:9} indicate that the electron density
increases toward the Galactic center, although the uncertainty is
large for the value inferred nearest to the Galactic center.  
These results also show that the inferred scale height of the gas,
which is computed from the fitted slope, increases toward the Galactic
center from H $\approx$ 190 pc at $R_G \sim 6.0\ $ kpc to H $\approx$
300 pc at $R_G \sim 4.1\ $kpc, although there is some overlap in the uncertainties.

Because the densities and scale heights derived above are
significantly different from those derived for the WIM near the Sun,
we hesitate to formally identify  the emission toward this window with
the WIM. This emission appears to characterize ionized gas that is
intermediate between the low density, large scale height WIM and the
so called Extended Low Density (ELD) \hii\ regions, which have
densities $n_e \approx 5-10\ \rm{cm}^{-3}$ and scale height $H \approx
100$ pc and are detected through radio observations \citep{Mezger78},
and Br-$\gamma$ recombination \citep{Kutyrev+03}.  Future studies that
characterize the properties of ionized gas in the inner Galaxy out to
much higher latitudes may help to clarify the nature of this emission
and its relationship to the WIM.  

\section{RELATIONSHIP BETWEEN IONIZED AND NEUTRAL GAS}
\label{sec:neutral}

The relationship between \hii\ and \hi, two principal states of
interstellar gas within the large-scale diffuse ISM, has yet to be
established.  
The \hii\ may be confined to the outer envelopes of \hi\ clouds
embedded in a low-density, hot ionized medium \citep{MO77};  the \hii\
may be the fully ionized component of a widespread warm neutral medium
\citep{MC93}; or the \hii\ may be well mixed with \hi\ in partially
ionized clouds \citep{SF93}.  
 
An examination of the \ha\ and \hi\ spectra toward the window, samples
of which are shown in Figure~\ref{fig:7}, offer a starting point for
such a study.  However, these lines of sight are close to the Galactic
plane, where much of the gas may be in molecular form, and where the
presence of multiple, interacting sources of ionization may be
influencing the dynamics and ionization of the gas in a complicated
way.  Furthermore, the relatively small size of the window ($\approx
30\ \rm{deg}^2$) compared to the angular resolution of the \ha\ and
\hi\ observations ($\sim 1\ \rm{deg}^2$) limits a morphological
comparison of the two phases. 
Nevertheless, some general spectral correlations seem to hold true in
this region and provide some insight in the relationship between the
\hi\ and \hii. 

From an examination of all of the \ha\ and \hi\ spectra in the
direction of the Scutum cloud, we find that in most cases,
kinematically distinct emission components from \hii\ appear near the
same velocity as the warm (broad) neutral \hi\ components, and that
these velocity components correspond to the Sagittarius and Scutum
spiral arms. 
However, there is little or no correlation between the
{\it{strengths}} of the \hii\ and warm \hi\ emission.  This is
consistent with the likely ionization mechanism of hydrogen.  If the
\hii\ is photoionized, then as long as the \hi\ is opaque to Lyman
continuum photons ($N_{\rm{H I}} \gtrsim 10^{18}$~cm$^{-2}$),  the
emission measure from \ha\ is related only to the intensity of the
ionizing flux and is thus unrelated to the total column density of
\hi.  
Interestingly, there are some directions in which very little high
velocity \hi\ emission is present, yet there is substantial \ha\
emission. An example of this is seen in the inner Galaxy near $\vlsr =
+100\ \kms$ in the three rightmost panels of Figure~\ref{fig:7}.   
This comparison would look even more dramatic if the 21 cm spectrum
were compared to extinction corrected \ha\ profiles, as shown in
Figure~\ref{fig:hacor}. 
These observations suggest that the ISM is substantially ionized in
these regions, with the \hi\ perhaps confined to small, unresolved
\hi\ clouds that populate the inner Galaxy \citep{Lockman02b}. 

We can compare the column densities of the ionized and neutral gas.
The midplane rms space density of electrons at different distances,
over 1 kpc intervals, is estimated in \S\ref{sec:scale} and shown in
Figure~\ref{fig:9}.  Multiplying these space densities by 1 kpc yields
midplane column densities $N_{H^+} = 1.7\phi^{1/2},\ 2.3\phi^{1/2},\
\rm{and}\ 3.3\phi^{1/2}\times10^{21}\ \rm{cm}^{-2}$ at heliocentric
distances $D_\odot = 3.0,\ 4.5,\ \rm{and}\ 6.0\ $kpc, respectively.   
Integrating the average \hi\ spectra at $b=0\dg$ between $22\dg < l <
28\dg$ from \citet{HIAtlas} over the same velocity/distance intervals,
yields midplane column densities $N_{HI} = 3.1,\ 2.9,\ \rm{and}\
3.5\times10^{21}\ \rm{cm}^{-2}$. 
We note that column densities of $H^{+}$ are upper limits, because its
volume filling fraction $\phi$ is less than 1.  In addition, the
column densities of \hi\ may be lower limits, because of potential
\hi\ self-absorption features in \hi\ spectra near the Galactic plane
\citep[e.g.][]{GibsonHISA+00}. 
Therefore, the fraction of gas that is ionized, $N_{H^+}/(N_{H^{+}} +
N_{HI})$, at each of these locations in the inner Galaxy are upper
limits.  
If $\phi$ and 21 cm absorption effects are the same at all distances,
then the relative amount of \hii\ increases toward the Galactic
center, with values of $N_{H^+}/(N_{H^{+}} + N_{HI})\ \rm{given\ by}\
0.35:0.44:0.49$ at $R_G \sim 6.0,\ 4.9,\ \rm{and}\ 4.1\ $kpc,
respectively. 

The extinction-corrected values of \iha\ are a direct measure of the
flux of Lyman continuum photons \citep[e.g.,][]{TRH98}. 
An \hi\ cloud illuminated on one side by a flux $F_{LC}$ of ionizing
photons emits one \ha\ photon for every 2.05 Lyman continuum photons
absorbed, provided the gas is optically thick to the Lyman continuum
photons and optically thin at \ha\ \citep{Osterbrock89}.   For a cloud
embedded in a uniform field of Lyman continuum radiation, the
intensity of \ha\ is increased by a factor of 2, because there are two
illuminated surfaces in projection.  Therefore \iha\ toward a
collection of $N_C$ clouds bathed in a Lyman continuum flux $F_{LC}$
is  given by 
\begin{equation}
\label{eq:10}
F_{LC} = 2.05\times10^6\ \frac{\iha}{2N_C}\ \rm{photons}\ \rm{cm}^{-2}\ s^{-1}
\end{equation}
with \iha\ measured in Rayleighs.
From the fits shown in Figure~\ref{fig:8}, our observations indicate
that the extinction corrected midplane \ha\ intensity is
$I_{\rm{H}\alpha}^0 \approx 130,\ 250,\ {\rm{and}}\ 490\ R$ at
heliocentric distances $D_\odot = 3.0,\ 4.5,\ \rm{and}\ 6.0\ $kpc,
respectively.  This corresponds to a one-sided Lyman continuum flux of
$F_{LC} = 1.3/N_C,\ 2.6/N_C,$\ and $5.0/N_C \times10^8$\ photons
cm$^{-2}$ s$^{-1}$.  If there is the same number of clouds present in
each 1 kpc interval over which \iha\ was measured, then the flux of
ionizing radiation increases by a factor of 4 from $R_G \sim 6.0\ $kpc
to $R_G \sim 4.1\ $kpc.

Thus both the Lyman continuum flux and the fractional amount of H$^+$
appear to increase toward the inner Galaxy.  This is consistent with
many observations that suggest an increase in star formation activity
in the Galaxy interior to the solar circle.

\section{RELATIONSHIP BETWEEN GAS AND DUST}
\label{sec:ebv}

\begin{figure}[htp]
\includegraphics*[angle=90,scale=0.9, trim= 0in 0in 0in -1in]{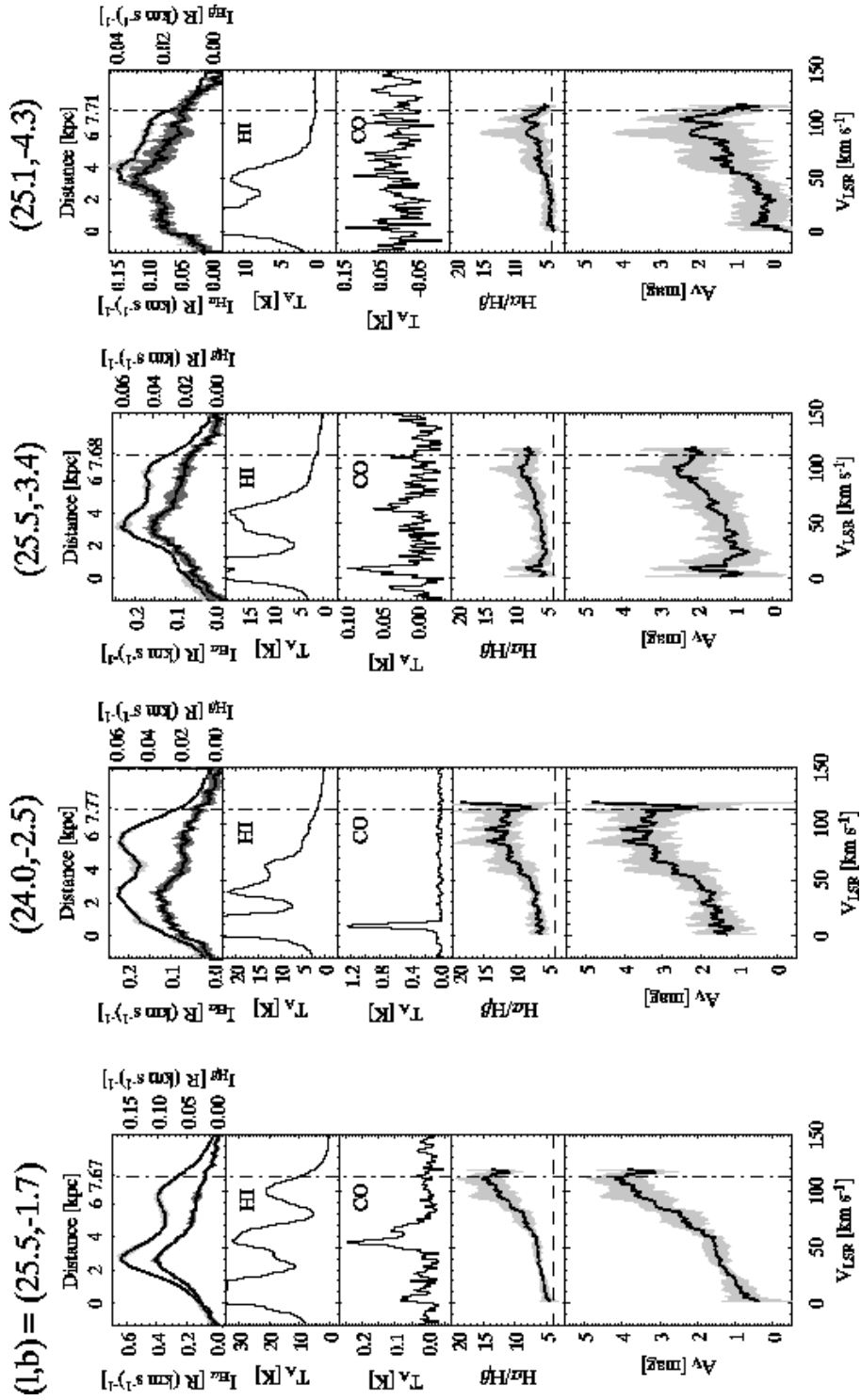}
\caption{Sample of spectra for four sightlines in this study, along
  with other tracers of the ISM. Notice the increase in \av\ with
  increasing distance/velocity, and the relationship between the shape
  of the \av\ curve and the \hi\ and CO spectra (see
  \S\ref{sec:ext_res}). \label{fig:7}} 
\end{figure}

In addition to the \ha\ and \hi\ spectra, a relationship appears to
exists between features in the molecular CO emission spectra and
values of \av\ derived from the optical \ha\ and \hb\ emission. 
Samples of these data are shown in Figure~\ref{fig:7}. As discussed in
\S\ref{sec:ext}, \av\ remains flat or increases with velocity
(distance) in all directions, consistent with the cumulative effects
of extinction with distance. 
In some cases, a sharp rise in \av\ is evident, indicative of the
presence of a dense concentration of dust.  
The leftmost panel in Figure~\ref{fig:7}, near $\vlsr = +60\ \kms$,
shows an example of this. 
Not suprisingly, we find strong CO emission at the same velocities as
this sharp rise in \av, confirming the well-known correlation between
the presence of molecular gas and interstellar dust.  
In other cases, such as in the second panel from the left in
Figure~\ref{fig:7}, \av\ rises smoothly in the absence of strong CO
emission, suggesting that interstellar dust is more smoothly
distributed  along the line of sight. 

These independent measurements of the extinction, when combined with
\hi\ and  CO spectra, allow us to quantify the relationship between
the dust  and gas content toward the window. 
\citet{BSD78} investigated the relationship between the color excess
\ebv\ and the total neutral hydrogen column density, \nh\ = $N_{HI} +
2N_{H_2}$, toward a population of several nearby ($d \lesssim 2\ $kpc)
hot stars. Their observations traversed sightlines with $\ebv < 0.5$,
from which they determined a ``definitive value" of \nh/\ebv $= 5.8
\times 10^{21}\ $atoms cm$^{-2}$ mag$^{-1} $. This quantity is cited
often and is used to estimate color excesses (and extinction
corrections) for objects in which \ebv\ cannot be determined
directly. 

\begin{figure}[htp]
\includegraphics[scale=0.75,trim=-2in 0in 0in 0in]{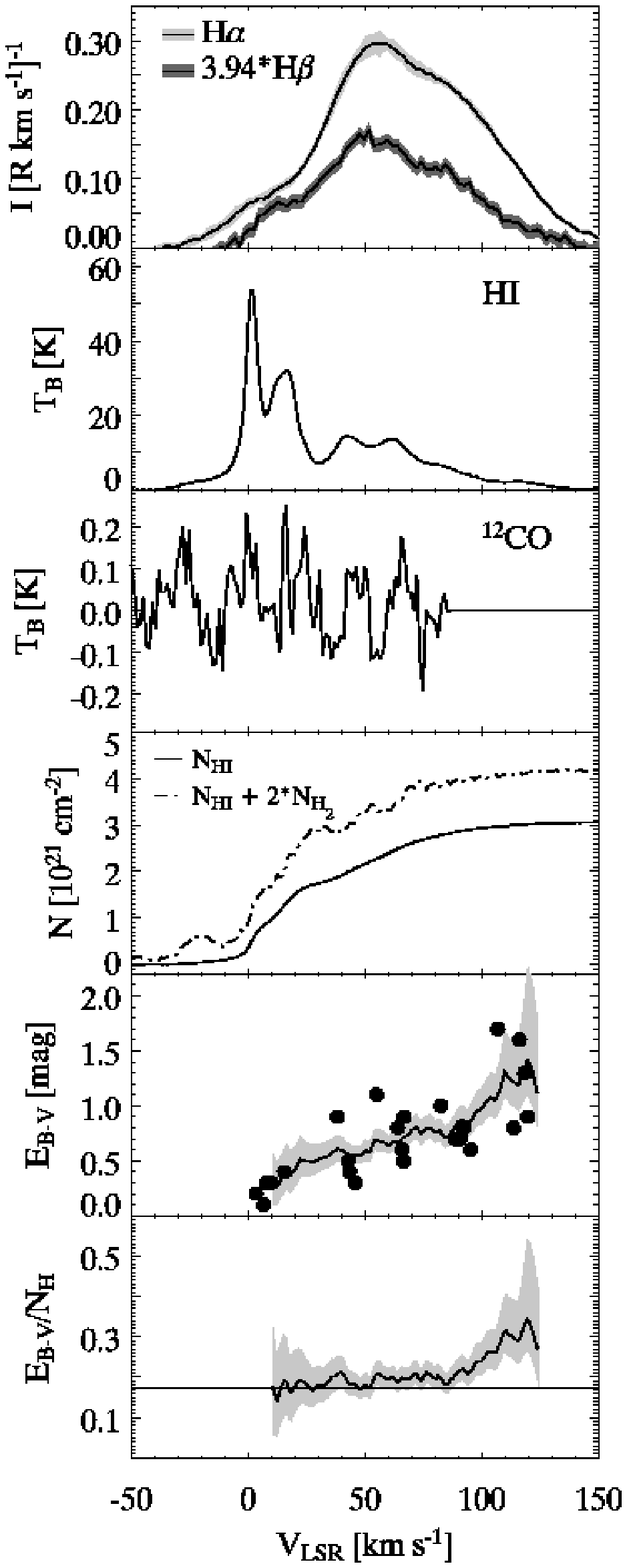}
\caption{ Comparison of \ebv/\nh\ toward the low extinction window
  from our data with the canonical value from \citet{BSD78}. The \hi\
  and CO data are from \citet{HIAtlas} and \citet{DHT01},
  respectively. The bottom panel, with \ebv/\nh\ given in units of mag
  per 10$^{21}$ cm$^{-2}$, shows good agreement between our data
  (\emph{shaded}) and the \citet{BSD78} value (\emph{straight line}),
  with evidence for a larger extinction per unit hydrogen atom in the
  inner Galaxy (see \S\ref{sec:ebv}).\label{fig:10}} 
\end{figure}

The lines of sight to the low extinction window contain substantially
more dust and gas than those studied in \citet{BSD78}.  However, the
same kind of analysis can be made, and it can be compared with their
canonical value. Any cumulative change in this value over a very long
distance can also be assessed. Figure~\ref{fig:10} summarizes this
analysis.  The top panel shows the average \ha\ and \hb\ spectra
toward five lines of sight toward the Scutum cloud. As with the
previous figures, the total uncertainty in the spectra are shown as
the shaded regions around the spectra, and the \hb\ spectra have been
multiplied by 3.94, the ratio of \iha/\ihb\ in the absence of
extinction.  
These five directions were chosen because of their low
uncertainty in \av, presence of high-velocity emission, and low CO emission
line strength.  This increased the accuracy of our determination of \nh/\ebv, and avoided directions where CO self-absorption may be present \citep{Minter+01, Jackson+02}.
The second and third panels show the average \hi\ and CO spectra within the WHAM beam in these directions, from the surveys of \citet{HIAtlas} and
\citet{DHT01}, respectively. 
The straight line at high velocities in
the CO spectra indicate a lack of data at those velocities.  The
fourth panel shows the cumulative column density of \hi, along with
\hi~+~2H$_2$, as a function of distance in units of $10^{21}$
cm$^{-2}$. This is calculated by integrating the \hi\ and CO spectra
from $\vlsr \approx -20\ \kms$ out to each velocity point, using a
CO-to-H$_2$ mass conversion factor of $X = 1.8\times 10^{20}\
\rm{cm}^{-2}\ (\rm{K}\ \rm{km}\ \rm{s}^{-1})^{-1}$\ from \citet{DHT01}. We
note that while the CO spectra appear quite noisy, there exists a
positive signal which significantly contributes to the total neutral
hydrogen column density at the higher velocities.  
The fifth panel shows the inferred values of the color excess \ebv\
along with its uncertainty.  The values of \ebv\ are only shown over
the velocity interval in which the uncertainties are low. The bottom
panel shows the ratio of \ebv/\nh, in units of mag per 10$^{21}$
cm$^{-2}$, along with a solid horizontal line at the location of the
canonical value from \citet{BSD78}. Note that this is the inverse of
the quantity \nh/\ebv\ discussed in \citet{BSD78}. 

Out to $\vlsr \approx +80\ \kms$, our derived value agrees with the
value from \citet{BSD78}. Beyond this velocity, there is an increase
in the amount of extinction per unit hydrogen column density, by about
a factor of two.  As seen in Figure~\ref{fig:10}, this is the result
of an increase in \ebv.  We note that the potential presence of very faint, broad \ha\ and \hb\ emission in the spectra would have been removed by our continuum subtraction process, particularly at the highest velocities. 
This effect would result in an underestimate of the \ha\ emission relative to \hb, and would increase the derived value of \ebv/\nh\ over its already enhanced value.
In addition, since we are measuring the {\it{cumulative}} extinction and column density, this increase suggests that the local value of this number in the inner Galaxy, at $D_\odot \gtrsim 5\ $kpc, is even higher than the observed value of $\approx$
0.3 mag per 10$^{21}$ cm$^{-2}$.   If a constant, elevated value is
present only over the last third of the distance out to the tangent
point, as the figure suggests, than the local value near the tangent
point is $\approx$ 0.6 mag per 10$^{21}$ cm$^{-2}$, more than 3 times
greater than the value near the Sun.

Recent work by \citet{LC05} have found the opposite trend toward regions with very low reddening.
Using high sensitivity \hi\ observations at high Galactic latitude, they find that \ebv/\nh\ $\approx$ 0.1 mag per 10$^{21}$ cm$^{-2}$,  which is almost half of the canonical value. 
Their data included sightlines with \hi\ column densities that are $\approx 10$ times below those included in \citet{BSD78}.
When combined with our results, it appears that the amount of reddening per unit hydrogen atom in the Galaxy is not a constant. Another parameter, perhaps related to the total \nh\ and/or Galactic environment, may be important in understanding the relationship between \ebv\ and \nh.

We note that although our derived values within 4 kpc of the Sun fall
near the canonical value, there are several uncertainties in this
analysis. 
Most importantly, \citet{BSD78} were able to directly
measure both \ebv\ through stellar photometry as well as \nh\ through UV absorption lines, whereas we have used \iha/\ihb\ as a proxy for
the extinction, and the emission lines from \hi\ and CO to estimate
the total neutral hydrogen column. 
To address some of the caveats regarding our estimates of the
extinction, particularly with the potential blending of the emission
components, we have modeled the average \ha\ and \hb\ spectra as a sum
of multiple Gaussian components, and compared the ratio of these
individual components to the smooth curve of \ebv. Each of the five
\ha\ spectra were fitted with five components, with fixed widths of
$25\ \kms$. The \hb\ spectra were fitted by fixing both the width and
location of the components, as determined from the fit to the
corresponding \ha\ spectra, since this emission comes from the same
atoms. The ratio of each of these components was converted into  \ebv,
and these are shown as dark circles in the fifth panel of
Figure~\ref{fig:10}.  This method yields ratios that are consistent
with the smooth, average curve, showing that the blending of the
emission lines is not an important effect. 
In addition, the relationship between \iha/\ihb\ and \ebv\ is not very
sensitive to potential changes in $R_V$ along these lines of sight
toward the inner Galaxy, as discussed in \S\ref{sec:ext}. 

Our calculation of \nh\ is also subject to some uncertainty.  The
intensity of CO spectra are near the level of the quoted uncertainty
in  the CO survey. However, when integrated over all velocities, the
integration yields a result several standard deviations away from the
noise.  The uncertainty in the cumulative CO emission is smallest at the highest velocities.
The subsequent conversion from this value
to an H$_2$ column density is not very well constrained. The
conversion factor, $X$,  is known to vary in different environments,
particularly in regions with non-solar metallicities, as is the case
in the inner Galaxy \citep{Wilson95, Simpson+04}.  The higher
metallicity in the inner Galaxy would result in a conversion that
overestimates the actual $N_{H_2}$ at the higher velocities, which in
turn would further increase our derived value of \ebv/\nh\ over its
already enhanced value.  
However, by selecting directions with low CO emission line strengths, we may have introduced a bias in our results which may not hold true for other sightlines in the inner Galaxy. 

If the dust properties in the {\it{ionized}} gas are similar to those in the neutral medium, then the increasing fraction of H$^+$ toward the inner Galaxy (\S6) would correspondingly increase \ebv/\nh.
However, the 40\% increase derived for $N_{H^+}/(N_{H^+} + N_{HI})$\
is far too small to account for a factor of three
increase in \ebv/\nh. 

The uncertainty in the beam filling factor of \hi\ and CO emission
within the 1\dg\ WHAM beam also may  bias our results. These
measurements were taken with a single dish telescope, and
therefore clumping of the emission into small knots can change the
interpretation of the data.  For example, consider a `two-phase'
medium, in which the column density of dust and neutral gas are bimodally distributed between two values in the beam.  
For a fixed average column density, this
clumpiness results in differentially more light passing through the
lower extinction regions, reducing the effective optical depth.  Our
analysis of \ebv/\nh\ has assumed that the radio emission is smoothly
distributed over some path length within the beam. If this is not
true, then our derived values of \ebv/\nh\ should be considered lower
limits, with this ratio elevated toward higher-density `clumps' within 
a 1\dg\ beam.  
The observed variation in the total column density
of CO and \hi\ among the five beams vary by a factor less than 2.  
If a density contrast of a factor of 2 between two `phases' is present within
a beam, then for an average total column density of $4 \times 10^{21}\ $ cm$^{-2}$, the effective \ebv/\nh\ is reduced by 10\% at most,
for all values of the relative areal filling factor of the two
phases. A reduction by 50\% is not reached until the density contrast
reaches 10, and then only for a limited range in areal filling
factors. We conclude that our values of \ebv/\nh\ are not strongly
affected by such systematic uncertainties. 

\section{PHYSICAL CONDITIONS OF THE IONIZED GAS}
\label{sec:ions}

Observations of the emission lines of \nii, \sii, and \oiii\ can be
used to explore the physical conditions of gas by comparing the
strength of these lines to that of \ha.  
For example, \nii/\ha\ traces variations in the electron temperature
of warm ionized gas, assuming that the gas is photoionized and that
that ionization fraction of N$^+$/N is not changing \citep{HRT99,
  RHT99}. The line ratio of \sii/\nii\ is nearly temperature
independent and can be used to estimate the ionization fraction of
S$^+$/S, while the ratio \oiii/\ha\ is a measure of the higher
ionization states of the gas. 
These lines have been used to characterize diffuse warm ionized gas within $\approx$ 2 kpc of the Sun \citep[e.g.,][]{HRT99}.  Observations of these lines toward the low extinction window provide a new and unique opportunity to examine
the potential changes in the physical conditions of the gas toward the
inner Galaxy.   

\begin{figure}[htp]
\includegraphics[scale=0.75]{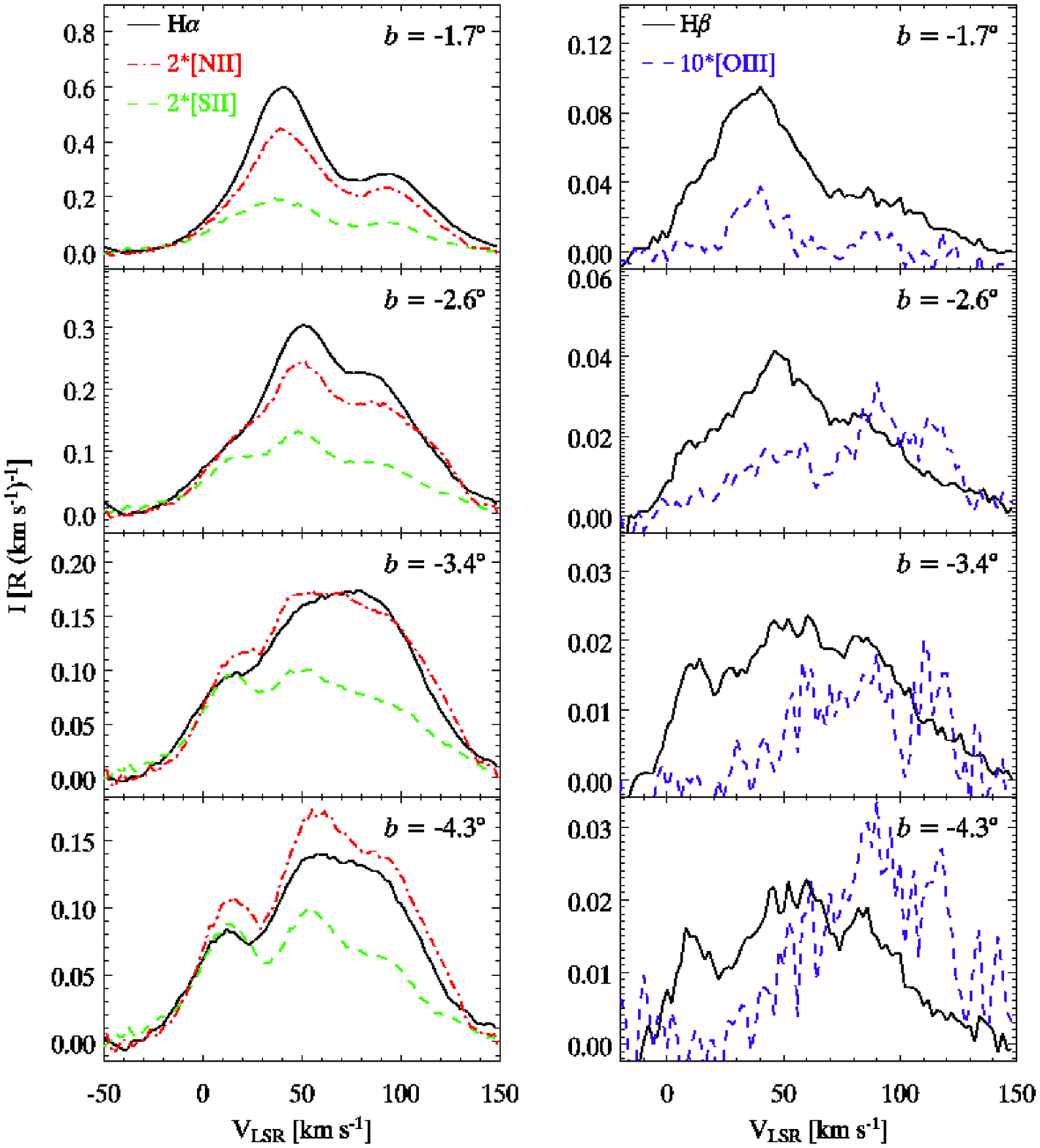}
\caption{ Emission line spectra toward the window. Thirteen sightlines
  were observed in emission lines that probe the temperature and
  ionization state of the gas. The left panel shows the average \ha\
  (\emph{black}), \nii\ (\emph{red}), and \sii\ (\emph{green}) spectra
  at each latitude.  The right panel is the same, but for \hb\
  (\emph{black}) and \oiii\ (\emph{purple}). Some of the spectra have
  been multiplied by the indicated values to facilitate a visual
  comparison of the profiles.  
Note the increase in \nii/\ha\ and \sii/\ha\ with increasing distance from the plane. 
Also note the large increase in \oiii/\hb\ with distance from the plane at high velocities (i.e., in the inner Galaxy).
\label{fig:11}}
\end{figure}

Results from some of our observations of these lines are shown in
Figure~\ref{fig:11}.  
Spectra taken at the same latitude within the window have been
averaged over longitude and appear in each of the panels in the
figure.  
The left panels shows the spectra of \ha\ (\emph{black}), \nii\
(\emph{red}), and \sii\ (\emph{green}) with increasing $|b|$, where
\nii\ and \sii\ have been multiplied by 2 to facilitate the comparison
of the relative line strengths.  
The right panel shows the spectra of \hb\ (\emph{black}) and \oiii\
(\emph{purple}), where \oiii\ has been multiplied by a factor of 10
for comparison. The similarity of wavelengths for spectra in each of
the two panels greatly reduces the effect of extinction on the
relative line strengths. 

There are several interesting trends in the data. While the \ha\
emission becomes weaker at higher $|b|$, both \nii/\ha\ and \sii/\ha\
increase. At a given latitude, \sii/\nii\ tends to decrease with
increasing distance (velocity). This effect is greatest for the
$b=-4.3\dg$ data.  
There are significant changes in \oiii/\hb\ with both latitude and
distance. At $b=-1.7\dg$, there are two velocity components in the
\oiii\ spectrum that correspond to components in the \hb\ spectrum,
with \oiii/\hb\ $\sim$\ 0.05 for both components. With increasing
distances from the plane, \oiii/\hb\ associated with gas in the inner
Galaxy (near the tangent point) increase by a factor of four or more,
from 0.05 at $b=-1.7\dg$ ($z=-180$ pc) to about 0.2 at $b=-4.3\dg$
($z=-450$ pc). 

A more quantitative description of these line ratios are shown in
Table~\ref{tab:1}. 
The spectra have been integrated over four discrete velocity
intervals, centered at +25, +50, +75, and +100 \kms, with a width of
25 \kms. The \ha\ and \hb\ data were used to estimate the extinction
\av\ as discussed above. The extinction at the wavelengths of the
other emission lines was calculated from \citet{CCM89}. All of the
data in the table have been corrected for extinction, and the line
ratios are given in energy units.  

We find that \nii/\ha\ increases from $\sim$\ 0.4 to $\sim$\ 0.6 as
$b$\ decreases from $-1.7\dg$\ to $-4.3\dg$, at all velocities. This
variation could be due to changes in the ionization state of N, the
abundance of N, or the temperature.  
If it is only temperature that is changing, then the data suggest that
$T_e$ increases from $\sim$\ 7000 K to $\sim$\ 8000 K with increasing
height above the plane, assuming N/H = 7.5$\times10^{-5}$ and N$^{+}$/N = 0.8 \citep{MCS97, Sembach+00}.
This increase in temperature in regions of lower \ha\ emission, or
lower density, is consistent with several similar observations
throughout the WIM \citep[][]{HRT99}.  
However, the abundance of N, relative to H, changes with distance from
the Galactic center \citep{ACW97}, with N/H greater by about a factor
of 1.8  in the inner Galaxy \citep{Simpson+04}. In this case there is
an increase in $T_e$ of about $\sim$\ 700 K from $b=-1.7\dg$ to
-4.3\dg, and the overall temperatures are lower by about 1000 K than
those quoted above. 

The interpretation of \sii/\nii\ also suffers from an uncertainty in
the assumptions about the physical state of the gas.  However, if
variations in \sii/\nii\ are due only to changes in the ionization
fraction of S$^+$/S (i.e., N$^+$/N is constant), then the data suggest
that S$^+$/S increases from $\sim$\ 0.25 to $\sim$ 0.4 with increasing
$|b|$ for the nearby gas, and is relatively constant  at $\sim$\ 0.25
in the inner Galaxy, assuming S/H = 1.86$\times10^{-5}$ \citep{AG89}. 

The \oiii\ data for the inner Galaxy show a definitive increase in
\oiii/\ha\ with increasing distance from the plane. This increase in
the inner Galaxy is much larger than the uncertainty in the line
ratios and is in contrast to other observations of the WIM in the
solar neighborhood that show \oiii/\ha\ {\it{decreasing}} with
distance above the plane (Madsen et al.~2005, in preparation). The
line ratios are all less than $\sim$ 0.06, which is consistent with
other observations in the diffuse WIM,  and is considerably less than
what is observed in traditional \hii\ regions. The increase in
\oiii/\ha\ with $|b|$ could be due to several effects, including an
increase in temperature or contributions from shock ionization (i.e.,
increase in the abundance of O$^{++}$).  
An increase in T by 1000 K (see above) would increase the \oiii/\ha\
ratio by about a factor of 2, which is much less than what is observed
(Table~\ref{tab:1}). This suggests that the O$^{++}$/H$^+$ ratio also
increases with increasing $|z|$ in the inner Galaxy, as has been
suggested by observations in other galaxies \citep[e.g.,][]{Rand97}.

\section{SUMMARY}
\label{sec:summary}

We have detected diffuse \ha\ emission from warm ionized gas toward
the inner Galaxy at velocities that place the gas near the tangent
point.  Observations of the Balmer decrement confirm that we are
seeing optical emission a few degrees away from the Galactic plane at
heliocentric distances greater than $\approx$\ 6 kpc.  
We have characterized the extinction in this region and identified a
$\sim$\ 5\dg$\times$5\dg\   low extinction window centered near \lb =
(27\dg, -3\dg) with a total \av\ $\sim$\ 3 out to the tangent point. 

This unique window into the inner Galaxy provides an opportunity to
investigate potential changes in the nature of the diffuse
interstellar medium in different Galactic environments using optical
nebular emission line diagnostic techniques.  
The \ha\ and \hb\ spectra have emission line profiles that are
consistent with emission from the local neighborhood, the Sagittarius
spiral arm, the Scutum spiral arm, and beyond.  
We have used a model for the rotation of the Galaxy to infer a
distance to these emission components, and to derive the distribution
and density of the gas. The rms midplane density and the scale height
increase toward the center of the Galaxy, with $\phi^{1/2}n_e^0$
$\approx$ 0.54, 0.75, and 1.1 cm$^{-3}$ and $H \approx$\ 190, 270, and
300 pc at $R_G \sim$\ 6.0, 4.9, and 4.0 kpc, respectively.  
A comparison between \ha\ and \hi\ emission in this window suggests
that the warm ionized gas may be associated with the surfaces of warm
neutral gas clouds. The \ha\ and \hi\ data also suggest that the flux
of Lyman continuum radiation increases by a factor of $\approx$\ 3 and
the fraction of hydrogen that is ionized increases by a factor of
$\sim$ 1.5 from $R_G = 6$\ kpc to 4 kpc toward the inner Galaxy.  
We have combined our extinction measurements with \hi\ and CO data
and found that within $\sim$\ 6 kpc of the Sun, the amount of
extinction per unit neutral hydrogen, \ebv/\nh\, is similar to the
canonical value of \citet{BSD78}. We also found evidence that this
ratio increases by a factor of $\approx$\ 2-3 at a Galactocentric
radius $R_G \lesssim$ 4 kpc, and suggest that \ebv/\nh\ varies with Galactic environment.  Observations of emission
from the other ions of \nii, \sii, and \oiii\ suggest that the
temperature and ionization state of the gas in the inner Galaxy
increases with increasing distance from the Galactic plane.  

Traditional studies of extinction in the Galaxy, through star counts
and stellar spectrophotometry, are focused toward directions of the
sky containing stars with reliable distances and sufficient
brightness, limiting these techniques to distances of about 2 kpc. The
method of using the Balmer emission line decrement from diffuse ionized gas,
presented here, has the advantage of being able to measure the
extinction out to much larger distances, provided there is strong
enough emission and/or low enough extinction, as in the case of the
Scutum Cloud. This approach is most reliable for patches of diffuse
emission at the largest distances, and is best suited for mapping out
extinction toward distant, low \av\ windows, complementing the locally
derived extinction from stellar data. 
This technique is also limited to directions in which the relationship
between velocity and distance can be reliably determined. Several
models for the distribution of \av\ in the Galaxy exist, and employ a
variety of observational and theoretical, and semi-empirical methods
over a range of optical and infrared wavelengths \citep{SFD98,
  Hakkila+97, DCL03}.  A large survey of \ha, \hb, and perhaps
Br-$\gamma$ near the Galactic plane, using the observational technique
described here, will help constrain these models and may lead to a
more accurate understanding of the three dimensional structure of gas
and dust in the Galaxy. 

The origin of this large window is not clear.  
The total neutral atomic hydrogen column density, as measured through the 21
cm emission line, is not substantially lower in this window compared
to regions around it, although \hi\ self-absorption and 21 cm emission
from beyond the tangent point must also be considered. Velocity
channel maps of \hi\ do not show evidence for a cavity or supershell,
as has been found in the Scutum supershell near \lb = (18\dg, -4\dg)
\citep{Callaway+00}. 
Emission from CO is elevated in some regions outside the window, consistent with the well-known correlation between the presence of optically obscuring dust and molecular gas.
However, diffuse infrared emission is neither enhanced nor suppressed substantially toward the window, as seen in maps from the IRAS and MSX instruments. There is no enhancement in the X-ray emission maps from ROSAT, although the large
columns of \hi\ would absorb most of the soft X-rays from the inner
Galaxy.
Future spectroscopic investigations toward this area of the Galaxy,
including deep spectrophotometry of distant stars, may help to
understand the origin of this window as well as the nature of the gas,
dust and stars in the inner Galaxy.

\section{ACKNOWLEDGEMENTS}

We thank the anonymous referee for a thorough review and helpful suggestions which improved the paper.
We are grateful to John Mathis for his insightful conversations about
the nature of interstellar dust, and to L.~Matt Haffner, Steve Tufte, and Kurt Jaehnig for their important roles in the continued success of WHAM. 
WHAM was built and continues to explore the warm ionized medium
through the generous support of the National Science Foundation
through grants AST96-19424 and AST02-04973. GJM acknowledges additional support
from the Wisconsin Space Grant Consortium. 



\clearpage

\newpage
\clearpage

\begin{deluxetable}{clllll}
\tablewidth{0pt}
\tablecaption{Summary of \nii, \sii, and \oiii\ Observations \label{tab:1}}
\tablehead{
\colhead{\vlsr\tablenotemark{a}} & \colhead{} & \multicolumn{4}{c}{$b$} \\
\cline{1-1} \cline{3-6} \phn & \phn & -1.7\dg & -2.6\dg & -3.4\dg & -4.3\dg}
\startdata
+25\kms     & \iha\ [R] &  22.8 &  11.9 &   8.8 &   6.4 \\
\phn        & \nii/\ha &  0.39 &  0.48 &  0.58 &  0.61 \\
\phn        & \sii/\ha &  0.19 &  0.28 &  0.38 &  0.42 \\
\phn        & \sii/\nii &  0.47 &  0.58 &  0.66 &  0.69 \\
\phn        & \oiii/\ha & 0.006 & 0.007 & 0.002 & 0.002 \\ \\
+50\kms & \iha\ [R] &  44.7 &  34.2 &  15.1 &   9.8 \\
\phn        & \nii/\ha & 0.37 &  0.40 &  0.54 &  0.61 \\
\phn        & \sii/\ha & 0.15 &  0.20 &  0.29 &  0.33 \\
\phn        & \sii/\nii &  0.41 &  0.50 &  0.54 &  0.54 \\ 
\phn        & \oiii/\ha & 0.010 & 0.013 & 0.018 & 0.042 \\ \\
 +75\kms & \iha\ [R] &  30.9 &  23.6 &  25.2 &  19.8 \\
\phn        & \nii/\ha &  0.41 &  0.39 &  0.48 &  0.56 \\
\phn        & \sii/\ha &  0.17 &  0.17 &  0.22 &  0.25 \\
\phn        & \sii/\nii &  0.42 &  0.44 &  0.45 &  0.45 \\ 
\phn        & \oiii/\ha & $<$0.000 & 0.010 & 0.018 & 0.024 \\ \\
+100\kms & \iha\ [R] &  56.3 &  44.6 &  27.4 &  19.8 \\
\phn          &\nii/\ha  &  0.40 &  0.44 &  0.49 &  0.57 \\
\phn          & \sii/\ha &  0.17 &  0.18 &  0.22 &  0.23 \\
\phn          & \sii/\nii &  0.42 &  0.40 &  0.43 &  0.40 \\
\phn        & \oiii/\ha & 0.002 & 0.018 & 0.037 & 0.065 \\
\enddata
\tablenotetext{a}{ Emission is averaged over longitude, and integrated
  over a 25 \kms\ interval centered at the given velocity. All of the
  data have been corrected for extinction, and the line ratios are
  given in energy units.} 
\end{deluxetable}

\end{document}